# A Survey on the Applications of Generative Artificial Intelligence in Automated Driving Systems Test Scenario Generation Methods

Ji Zhou (1), Yongqi Zhao (1), Yixian Hu, Hexuan Li, Zhengguo Gu (1), Nan Xu (2), Arno Eichberger (1)

(1) Institute of Automotive Engineering, Graz University of Technology, Graz, Austria, (2) National Key Laboratory of Automotive Chassis Integration and Bionics, Jilin university)

**Abstract**

Ensuring the safety and reliability of Automated Driving Systems (ADS) remains a critical challenge, as traditional verification methods such as large-scale on-road testing are prohibitively costly and time-consuming. To address this, Scenario-Based Testing has emerged as a scalable and efficient alternative, yet existing surveys provide only partial coverage of recent methodological and technological advances. This review systematically analyzes 31 primary studies, and 10 surveys identified through a comprehensive search spanning 2015–2025; however, the in-depth methodological synthesis and comparative evaluation focus primarily on recent frameworks (2023–2025), reflecting the surge of Artificial Intelligence (AI)-Assisted and multimodal approaches in this period. Traditional approaches rely on expert knowledge, ontologies, and naturalistic driving or accident data, while recent developments leverage generative models, including large language models, generative adversarial networks, diffusion models, and reinforcement learning frameworks, to synthesize diverse and safety-critical scenarios. Our synthesis identifies three persistent gaps: the absence of standardized evaluation metrics, limited integration of ethical and human factors, and insufficient coverage of multimodal and Operational Design Domain (ODD) –specific scenarios. To address these challenges, this review contributes a refined AI-specific taxonomy with multimodal extensions, an ethical and safety checklist for responsible scenario design, and an ODD coverage map with a scenario-difficulty schema to enable transparent benchmarking. Collectively, these contributions provide methodological clarity for researchers and practical guidance for industry, supporting reproducible evaluation and accelerating the safe deployment of higher-level ADS.

## Abbreviations

SBT Scenario-Based Testing
ADS Automated Driving Systems
AI Artificial Intelligence
ODD Operational Design Domain
SAE Society of Automotive Engineers
SOTIF Safety of the Intended Functionality
ADAS Advanced Driver Assistance Systems
HD High Definition
RGB Red Green Blue
LiDAR Light Detection and Ranging
LLM Large Language Model
GAN Generative Adversarial Network
DM Diffusion Model
RNN Recurrent Neural Network
VAE Variational Autoencoder
VLM Vision-Language Model
RL Reinforcement Learning
AST Abstract Syntax Tree
AII Academic Influence Index
RAS Resource Accessibility Score
OCS ODD Coverage Score
NDD Naturalistic Driving Data
ITTC Inverse Time to Collision
CT Combinatorial Testing
ASAM Association for Standardization of Automation and Measuring Systems
NLP Natural Language Processing
RAG Retrieval Augmented Generation
BEV Bird's Eye View
AV Automated Vehicle
AD Autonomous Driving
TTC Time-To-Collision

VRU Vulnerable Road User

# 1 Introduction

Scenario-Based Testing (SBT) has emerged as a pragmatic alternative to large-scale on-road validation [1], which is impractical for SAE L2-L5 automation due to prohibitive evidence requirements. Standards such as Safety of The Intended Functionality (SOTIF) [2] and UN/ECE R157 [3] have formalized scenario-centric evaluation to target representative and safety-critical situations within an ADS's ODD. In this review, we synthesize methods that operationalize SBT through rule-based, data-driven, and AI-assisted generation, and we assess their capacity to produce realistic and diverse scenarios at scale.

The development of scenario generation techniques has evolved with ADS complexity and standardization. Before 2017, validation relied on empirical methods like expert-defined test cases [4]. Since 2015, this field has witnessed transformative advancements driven by several methodological innovations, standardization efforts, and the integration of generative AI.

Early validation of ADS was historically built upon methodologies originally developed for Advanced Driver Assistance Systems (ADAS). In the ADAS era (SAE L0-2 [5]), testing relied predominantly on large-scale on-road driving, and comprehensive frameworks such as the one proposed by Stellet et al. [6] provided the foundation for systematic evaluation. However, as ADS (SAE L3-5 [5]) introduced far more complex functionalities and assumed responsibility for the entire dynamic driving task within defined ODDs, the limitations of pure on-road testing became evident. Kalra and Paddock [7] demonstrated in 2016 that proving statistical safety for higher-level ADS would require more than miles of real-world driving—an impractical requirement in terms of both cost and time. This recognition triggered the paradigm shift in 2017 toward Scenario-Based Testing (SBT), where representative and safety-critical traffic scenarios are systematically extracted from real-world datasets to enable scalable and reproducible validation [8].

In 2019, the PEGASUS project formalized Scenario-Based Testing as a standardized methodology for ADS safety assurance [9]. Building on this foundation, Feng et al. [10] proposed intelligent driving tests that combined naturalistic and adversarial environment generation, demonstrating how systematic scenario extraction could enhance both coverage and criticality. Later, the SOTIF Standard (2022) revision reports expanded guidance to V2X communication scenarios [11]. Concurrently, large-scale datasets like the Waymo Open Motion Dataset [12] provided interactive driving data, enabling the development of data-driven scenario generators. The 2023 debut of GAIA-1 reports photorealistic multi-ADS simulation modal video synthesis from text or sensor inputs and has been cited as a promising direction for scenario synthesis [13]. This breakthrough was followed by specialized frameworks: SimNet (2021) learned reactive simulation agents directly from real-world observations [14]. TrafficGen (2023) leveraged graph-based Variational Auto-Encoders (VAEs) to model interactive traffic participant behaviors [15], while SEAL (2024) combined RL with human driving skills to generate safety-critical scenarios [16]. Recent advancements in 2025 have focused on unified multimodal generation. Limited to September 2025, UMGen [17] introduced a parallel autoregressive framework capable of generating ego-vehicle motions, HD maps, and traffic participants in a single model, while Genesis [18] presents empirical results suggesting cross-modal consistency between RGB videos and LiDAR point clouds using shared latent spaces. These models aim to improve scenario realism and diversity, with early evidence suggesting potential for closed-loop ADS simulation. However, broader validation remains ongoing.

Two key challenges remain: ensuring scenario diversity—especially for edge cases, which suffer from what Liu and Feng [19] characterize as the "curse of rarity," where safety-critical events are inherently underrepresented in naturalistic datasets and thus difficult to capture or validate at scale—and integrating ethical considerations into generated scenarios. The integration of LLMs for text-driven scenario specification, as demonstrated by preprint reports such as Txt2Sce [20], also indicate potential for bridging human intent and machine-executable test cases; these claims require independent validation. As ADS is in transition to industrial standard like L2+ [21] and SAE standard defined L3, L4 or L5 [5] automation, especially in L2+ adaptation —with more abilities than L2 function (e.g. Hands-Off in short duration) yet with significant needs of SBT to advance. SBT and its generation will play an increasingly critical role in large-scale safety verification.

# 2 Paper Organization

This survey is organized as follows. Section 3 presents the literature search process and summarizes the main contributions of this review. Section 4 introduces the foundations of Scenario-Based Testing for ADS together with source data collection and preparation. Section 5 reviews traditional and AI-assisted scenario generation methods. Section 6 presents the comparative evaluation framework and applies the proposed metrics to representative methods. Section 7 discusses the broader methodological implications, including multimodal integration, ethical considerations, ODD coverage, and geographic trends. Section 8 outlines the limitations of this review and directions for future work, and Section 9 concludes the paper. The appendices provide supporting terminology, checklists, scoring rubrics, and the scenario-difficulty schema.

**Fig.**1 presents the overall organization of this review. Rather than detailing individual model families, the figure highlights the main analytical layers of the paper: literature review and methodological framing, source data collection and preparation, scenario generation methods, and evaluation and synthesis. Within the generation layer, the review distinguishes between traditional approaches and

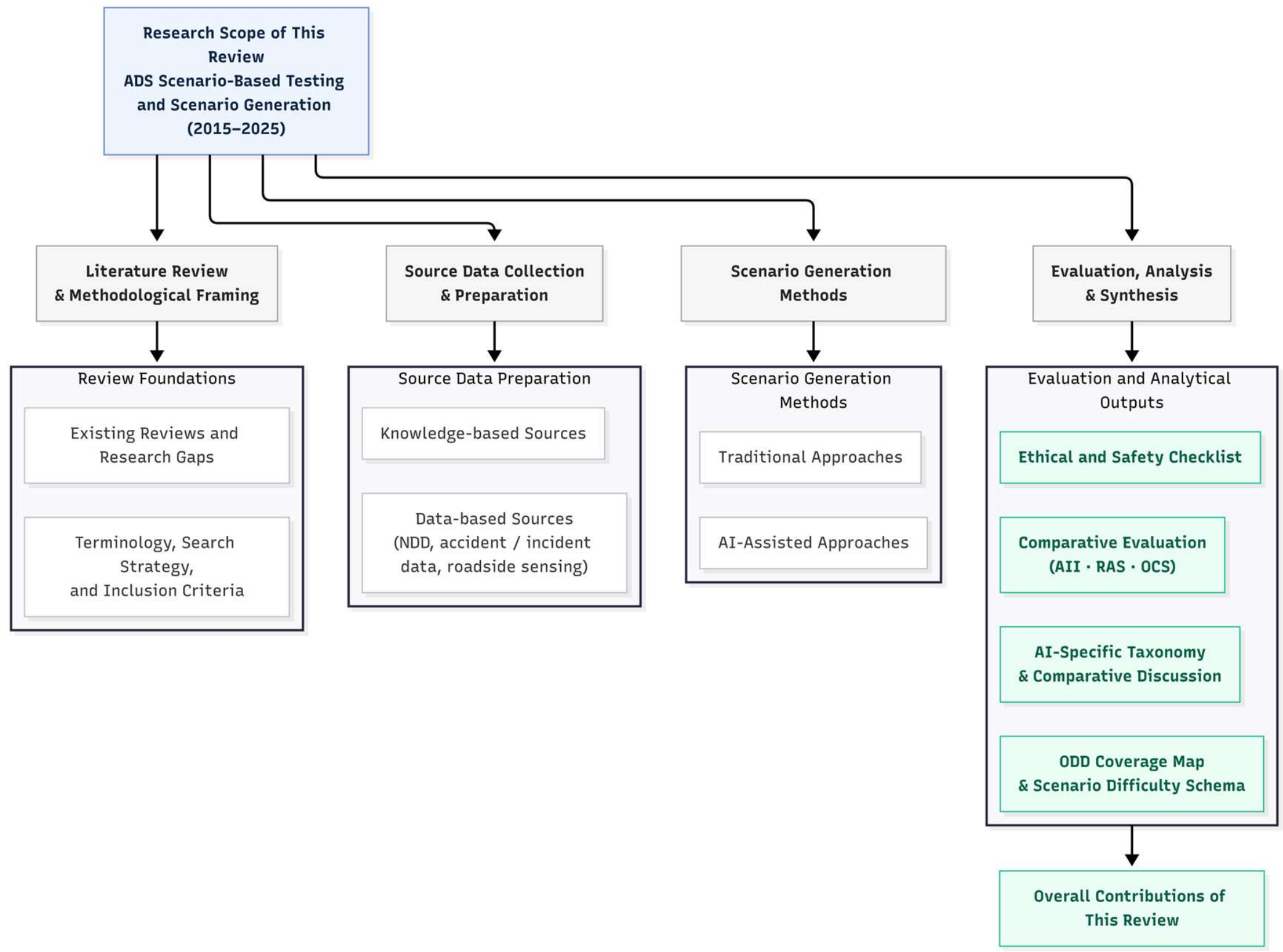

**Fig.**1 The Research Organization of This Review

AI-assisted approaches. The AI-assisted workflow and the detailed taxonomy of recent AI-assisted methods are discussed separately in Section 5.2 and illustrated in Figs. 4 and 5, respectively. The final layer summarizes the main analytical outputs of the review, including the metric suite, ethical and safety checklist, and the ODD coverage and scenario-difficulty framework.

# 3 Methodology and Literature Review

## 3.1 Search Criteria

This section outlines the paper query methodology employed in this study, encompassing paper selection criteria, data sources, and search keywords.

### 3.1.1 Inclusion Criteria

Primary studies focus on Scenario-Based Testing and test-scenario generation for ADS and Autonomous Driving.

### 3.1.2 Exclusion Criteria

a. Other safety assessments
(e.g., formal verification, attack)
b. Irrelevant publications

### 3.1.3 Timeframe and Publication Status

The literature search covered the period from January 2015 up to October 2025. Both peer-reviewed publications and preprints (e.g., arXiv) were included, providing that they contained sufficient methodological detail relevant to scenario generation for ADS. Preprints were explicitly marked in the reference list to distinguish them from formally published works. This policy ensures comprehensive coverage of the rapidly evolving field while maintaining transparency about the publication status of the cited studies.

While the paper search covered 2015–2025 to ensure completeness, the comparative analysis and proposed contributions emphasize developments from 2023 onward, as these represent the most significant methodological shifts toward AI-assisted scenario generation.

### 3.1.4 Keywords

The keywords used in the search are listed below:

a. Scenario-Based Testing (SBT) **(Mandatory)**
b. Automated Driving Systems (ADS)
c. Advanced Driver Assistance Systems (ADAS)
d. Scenario Generation

To ensure comprehensive coverage, Boolean operators were applied to combine mandatory and optional keywords. The core query string was constructed as:

```
("Scenario-Based Testing" OR "Scenario Generation")

AND ("Automated Driving Systems" OR "Autonomous Driving" OR
"Automated Vehicles" OR "Self-driving Cars")

OR ("Advanced Driver Assistance Systems" OR "ADAS")
```

In addition, synonym expansion was employed to capture variations in terminology across different studies. For example, *Automated Driving Systems* were extended to include *Autonomous Driving*, *Automated Vehicles*, and *Self-driving*, while *Scenario-Based Testing* was paired with *Scenario Generation* and *Scenario Extraction*. This approach reduced the risk of omitting relevant literature due to inconsistent terminology usage among authors.

## 3.2 Search Result Overview

A comprehensive literature search was conducted across four electronic databases (Google Scholar, Web of Science, Scopus, and IEEE Xplore) covering publications from 2015 to 2025. The initial search yielded about 1,300 potential articles. After removing 250 duplicates, 1050 articles underwent title and abstract screening. Based on predefined inclusion and exclusion criteria, 928 articles were excluded at this stage. The remaining 122 articles were retrieved for full-text assessment, of which 11 were further excluded due to insufficient methodological detail or irrelevance to scenario generation for automated driving systems.

While these span 2015–2025, most frameworks analyzed in depth were published in 2023–2025, reflecting the surge of AI-assisted scenario generation. Ultimately, 31 articles and 10 relevant surveys met the inclusion criteria and were included in this review. The detailed selection process is illustrated in **Fig.** 2.

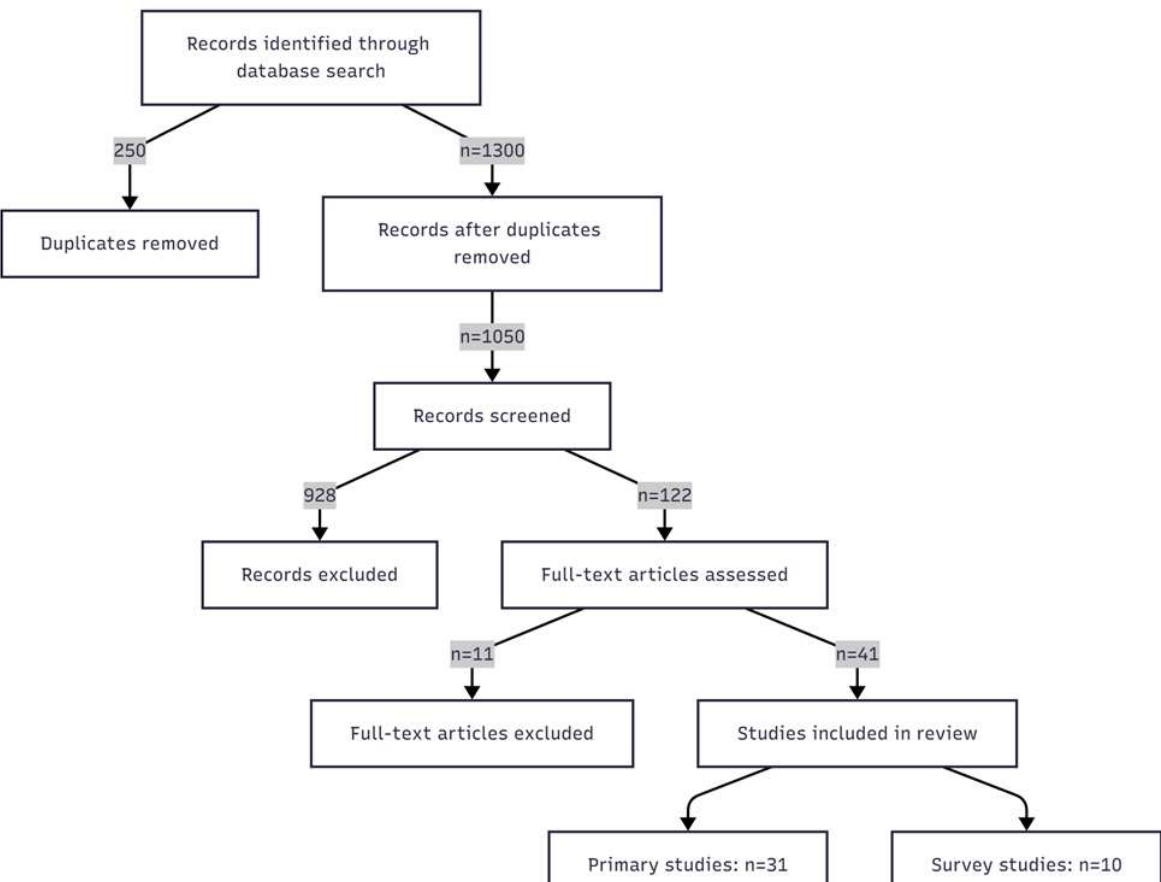

**Fig.** 2. Flow Diagram of Literature Selection

### 3.2.1 Existing Reviews

Sun et al. [22] made a valuable contribution by systematizing automation pipelines for Scenario-Based Testing, providing an early foundation for reproducibility. However, their limited coverage of data-driven generation techniques means that rare but safety-critical events—such as sudden pedestrian emergence or multi-agent conflicts—may not be adequately represented. This gap directly motivates the methodological framing adopted in this review, including an AI-specific taxonomy with multimodal extensions and complementary discussion of data-driven and AI-assisted approaches. While Cai et al. [23] provided a structured taxonomy of data-driven approaches, their analysis overlooked the methodological risks of dataset bias and limited generalizability across diverse ODDs. Ding et al. [24] advanced the field by offering a rigorous taxonomy for safety-critical scenarios, which remains a useful reference for structured evaluation. Yet, because their taxonomy does not incorporate cross-modal integration (e.g., RGB–LiDAR fusion), it risks becoming static in the face of evolving generative pipelines. This limitation highlights the need for our ODD coverage map and scenario-difficulty schema, which are designed to remain adaptable as new modalities and interaction complexities emerge. Schütt et al. [25] provided comprehensive catalogues of scenario generation methods, with detailed technical evaluation criteria such as coverage and diversity. However, their framework does not explicitly integrate human-in-the-loop validation or ethical safeguards. This omission may inadvertently allow technically efficient but ethically incomplete methods to be adopted, for example, by underrepresenting vulnerable road users or overlooking privacy-sensitive data. Cai et al. updated scenario generation strategies but excluded foundation model applications [26]. This lag in incorporating foundation models reflects a methodological inertia that may hinder timely adaptation to paradigm shifts in generative AI. Furthermore, while Feng et al. [27] proposed adaptive frameworks for testing scenario library generation in connected and automated vehicles, demonstrating how systematic library construction can balance representativeness and criticality, existing reviews have not yet synthesized these library-based approaches with the recent surge of generative AI methods, leaving a methodological gap between structured test catalogs and emergent synthesis techniques. Yang et al. [28] explored LLM integration in autonomous driving yet provided limited analysis of scenario testing applications. Their survey remained largely conceptual, without addressing the methodological gap between natural language specifications and machine-executable test cases. Zhao et al. [29] offered a timely survey on LLM-based scenario testing, highlighting the potential of natural language interfaces for scenario specification. Nevertheless, their mono-modal focus excludes multimodal models such as GAIA-1 [13] or Genesis [18], which are designed to ensure cross-modal consistency between text, video, and LiDAR. Without considering such models, LLM-generated scenarios may remain linguistically coherent but fail to capture physical constraints, for example, generating trajectories that violate kinematic feasibility. Gao et al. [30] surveyed foundation models in scenario generation but lacked discussion on L4/L5 validation metrics. Their omission of uncertainty-aware evaluation undermines the methodological robustness of their survey. Celik et al. [31] reviewed LLM-driven test case generation without extending to complex scenario synthesis. This narrow task orientation limits the methodological generalizability of their findings.

This rapid evolution has continued beyond the cutoff of earlier surveys. In particular, recent review efforts have begun to focus explicitly on foundation-model-based scenario generation and scenario analysis, while individual studies have also demonstrated that LLMs can now be used to translate natural-language descriptions into executable ADS test scenarios with substantially reduced manual configuration effort [30], [32]. These developments reinforce the need for the present review to maintain a strong emphasis on recent AI-assisted and multimodal methods rather than relying solely on earlier taxonomic distinctions.

#### 3.2.2 Consensus Findings across Studies

Several consistent findings emerged from the synthesis of included literature reviews:

a. **Scenario Taxonomy Development**

Seven studies proposed scenario classification systems, with the most common categories being functional scenarios, logical scenarios, and concrete scenarios.

b. **Validation Challenges**

All studies identified scenario validation as a major challenge, citing the lack of standardized metrics for scenario quality and relevance.

c. **Criticality Assessment**

Four studies have developed methods for identifying safety-critical scenarios, predominantly focusing on collision avoidance and edge cases.

d. **Data Scarcity**

Seven studies emphasized the scarcity of high-quality scenario data, particularly for rare but critical events.

### 3.3 Research Gaps and Contributions

Existing surveys reveal persistent methodological limitations, including the absence of multimodal advances, standardized evaluation metrics, insufficient integration of human and ethical factors, and inadequate coverage of multimodal and ODD-specific scenarios. These gaps hinder reproducibility, fairness, and comprehensive scenario diversity in ADS testing. Building on these observations, this review provides four key contributions:

#### 3.3.1 Comprehensive Methodological Synthesis Focused on the 2023–2025 Inflection.

This review consolidates scenario generation methods from 2015–2025, with particular emphasis on the 2023–2025 period during which AI-assisted and multimodal approaches became dominant. To support clearer comparison among recent frameworks, we introduce an AI-specific multi-axis taxonomy that interprets methods by their generation strategy, learning paradigm, modality coverage, and controllability, while retaining model-oriented categories as practical entry points for discussion.

#### 3.3.2 A Reproducible Three-Axis Metrics System and its Application.

We propose a unified metric suite (AII, RAS, OCS) —AII for scholarly impact, RAS for reproducibility, and OCS for scenario coverage—providing formal definitions and applying them to recent frameworks for transparent benchmarking.

#### 3.3.3 Operational Ethical & Safety Checklist for Scenario Generation.

We provide a concise checklist translating ethical principles—such as bias mitigation, privacy, and standards alignment—into auditable criteria for scenario generation.

#### 3.3.4 ODD Coverage Mapping and Scenario-Difficulty Schema.

We introduce an ODD coverage map and a scenario-difficulty schema to standardize reporting and classify scenarios by complexity and risk, enabling consistent benchmarking across frameworks.

Collectively, these contributions establish a structured foundation for reproducible evaluation and ethically aligned scenario generation in ADS validation.

## 4 Overview of SBT and Scenario Generation Preparation

### 4.1 SBT for ADS

The creation of test scenarios for Automated Driving Systems (ADS) is intended to recognize and derive specific situations from all kinds of traffic environments, with the goal of evaluating the performance and robustness of ADS in complex and dynamic road settings. Methods for extracting test scenarios must consider the variety of road users (such as vehicles, pedestrians, and bicycles), dynamic road conditions, and potential extreme occurrences like sudden pedestrian crossings or road obstacles. The variability of these factors makes test scenario generation a vital component of the ADS development process [24].

### 4.2 Generation Workflow

**Fig.**3 illustrates the generic workflow of test scenario generation in this review. The process begins with source data collection, which integrates both knowledge-based inputs, such as rules, ontologies, regulations and expert knowledge, and data-based inputs, including naturalistic driving data, accident records, and multimodal sensor observations. These inputs are subsequently curated and transformed into a scenario database or other structured representations that support downstream generation.

From this prepared input space, scenario generation proceeds through two broad methodological routes. Traditional methods rely on knowledge-driven derivation and data-driven extraction, typically using parameterization, classification, logical reasoning, or structured diversification to derive concrete scenarios. AI-assisted methods, by contrast, generate scenarios through end-to-end or hybrid pipelines that exploit learned representations and multimodal transformations. Both routes ultimately produce concrete scenarios in forms such as executable scripts, trajectories, or perceptual scene outputs, which are then subjected to evaluation and validation. In this way, **Fig.**3 emphasizes the shared workflow logic of scenario generation at a generic level, while the AI-assisted workflow and the corresponding methodological taxonomy

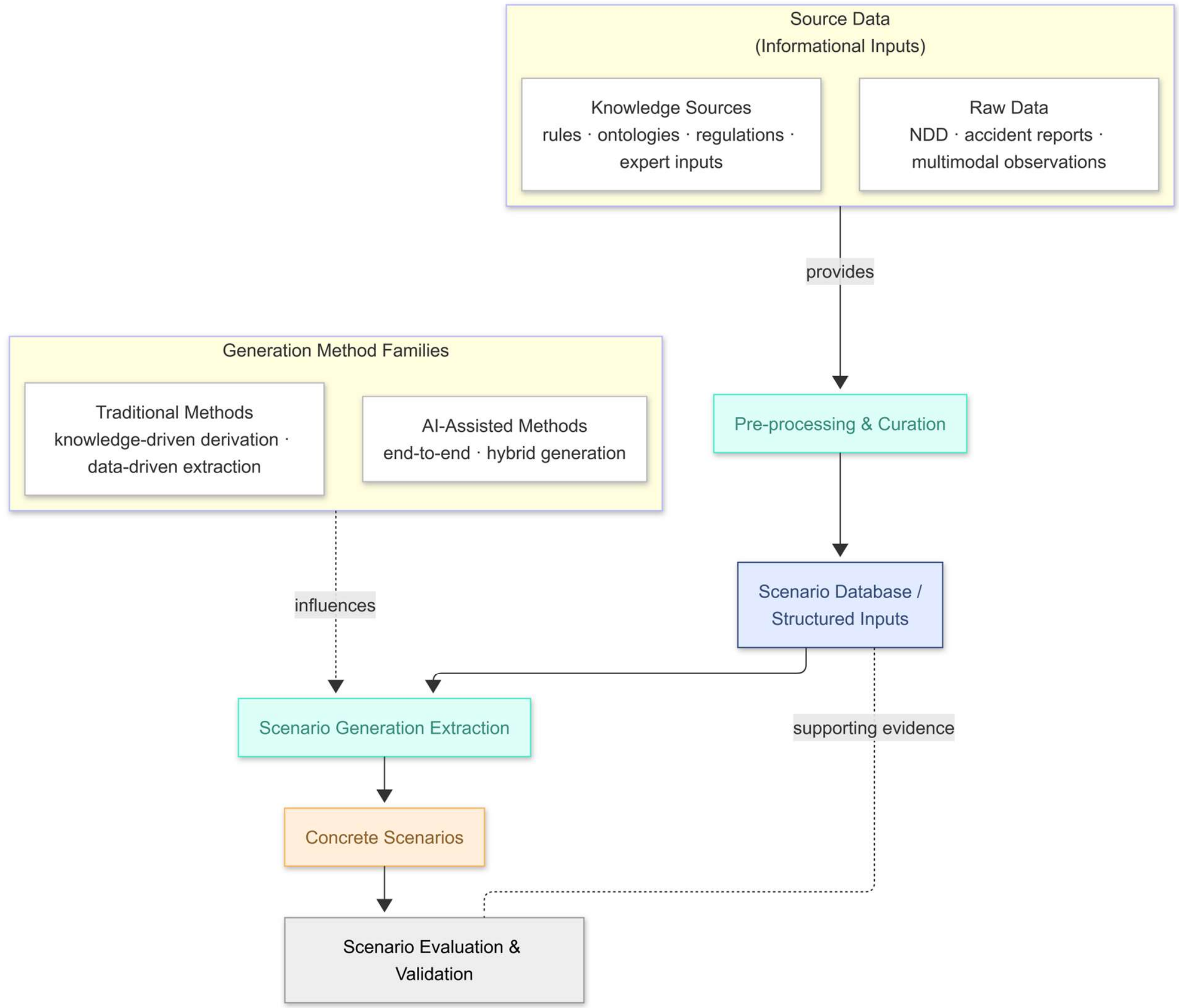

**Fig.**3 Generic Test Scenario Generation Process

are presented separately in Section 5.2 and illustrated in Fig. 4 and Fig. 5.

## 4.3 Source Data Collection

The validity and comprehensiveness of source data are vital for the quality of a scenario database, and further, for the generation of optimal scenarios. Many studies utilize data from the physical world. Meanwhile, some research adopts simulation-based approaches to obtain source data. This section presents approaches to acquiring such data, with several conclusions provided at the end.

### 4.3.1 Knowledge-based

In the context of Scenario-Based Testing for Automated Driving Systems (ADS), knowledge-based data collection methods are critical for generating relevant and realistic test scenarios, especially in structured environments like highways or city centers where predefined conditions are prevalent [33], [34], [35], [36]. These methods leverage various predefined information sources to identify and gather raw data for scenario generation, transforming potentially unstructured and ambiguous initial data into a computationally analyzable and human-understandable format.

One prominent approach in knowledge-based data collection involves the utilization of ontological concepts, which are advanced knowledge formalization tools [37], [38]. Ontologies provide a structured representation of knowledge by defining concepts and their relationships within a particular domain [39]. For ADS testing, this means defining elements such as road types, traffic rules, vehicle behaviors, and environmental conditions in a machine-understandable way [33]. This structured approach helps in extracting usable and clear rules from raw knowledge, leading to the generation of explainable scenarios that are interpretable by both humans and machines. For instance, ontology-based knowledge can provide a machine-understandable representation of roads, intersections, traffic rules, and driving behaviors, which is essential for driving simulations and traffic light optimization.

Another key method is the integration of expert recommendations and existing traffic regulations. This involves formalizing human expertise and their ability to make knowledge-based decisions, often through the systematic derivation of causal Bayesian networks based on ontologies [34]. Experts can abstract and transfer knowledge to new situations, making their input invaluable for defining what constitutes a typical or critical scenario. Data collection can also involve in-depth interviews and surveys to capture stakeholders' preferences and expert insights, which are then used to define and select criteria for scenario generation and evaluation [35].

#### 4.3.2 Data-based

Data-driven solutions can be categorized into two parts based on their sources: naturalistic driving data and accident-based data. We will discuss the differences and similarities between these two sources separately.

**Naturalistic Driving Data (NDD)**

This kind of data is derived from actual vehicles operating on real road networks. NDD collection primarily relies on two methods: vehicle-based and fixed-sensor-based approaches.

**Vehicle-based methods**

Which constitute the majority of publicly available driving datasets [23], primarily rely on onboard sensors such as LiDAR and cameras to record data, which is then fused during post-processing to generate high-quality usable datasets.

**Fixed-sensor-based method**

It is moreover primarily achieved through sensors fixed along the roadside or drones, which are significantly less costly than the former while still providing detailed trajectory information. The Ko-PER dataset [36] demonstrates driving data collected by installing LiDAR and cameras at public transit intersections in Germany. The highD dataset [40] proposed also comprises vehicle trajectories on German highways collected via drones. Bock et al. [41] achieved pixel-level accuracy in extracting vehicle and pedestrian trajectories using deep learning algorithms.

These traffic scenarios can be naturally extracted from real-world driving data. However, given the virtually limitless driving situations that may arise, the volume of real-world driving data required for collection and use in autonomous vehicle testing is enormous. However, collecting NDD using special vehicles demands significant investments, whereas fixed sensors are limited to monitoring a specific area. Furthermore, since many large-scale NDD datasets are not publicly accessible, we can only attempt to reduce the substantial demand for raw data during generation to satisfy the practical testing requirements for raw data. Reference [42] mentioned a few challenges it faces:

a. Limited quantity and variety.
b. Expensive and time-consuming.

Recent infrastructure-centered work further extends the value of fixed-sensor data by demonstrating that roadside perception can operate at network scale rather than only at isolated intersections or short road segments. A representative 2025 framework combines large-scale roadside sensing with distributed multi-access edge computing, enabling continuous tracking of tens of thousands of vehicles across a 157 km highway network instrumented with 1,777 sensors [43]. Reported performance includes average longitudinal and lateral errors of 2.14 m and 0.84 m, an average speed error of 1.91 kph, and processing latency below 340 ms [43]. For scenario generation research, this line of work is important because it suggests a path toward richer V2X-supported traffic observations that can capture broader spatial interaction structures and denser traffic flow patterns than those available from single-camera or drone-based datasets.

**Accident Data**

In addition to raw data from actual driving, some enterprises, academic institutions, or organizations are also utilizing non-driving, non-experiential data sources for scenario generation. Like the "Pre-Crash Scenario Typology" from NHTSA [44], it provides accident analysis from a statistical perspective while also offering a basic description of the accident scenario and accompanying illustrations.

Additionally, CIREN and NMVCCS also provided more information related to accidents. R. Queiroz et al [45] utilized these textual scenario descriptions as Waymo does, and recreated 18 usable concrete scenarios. Gambi et al. [46] proposed their semi-automated approach for generating simulations: CRISCE framework which mainly contributes to extract information from accident sketches then generate a high fidelity simulated scenario according to it.

From a comparative perspective, one of the advantages of accident data is its low cost and minimal labor requirements. Cai et al. [32] proposed a framework that is able to extract informational scenario from it, achieving the fully automatic raw-data (in natural language) processing and concrete scenario generation.

### 4.4 Summary of Source Data Collection

This section aims to provide valid and comprehensive source data for generating test scenarios in Automated Driving Systems (ADS) and constructing a high-quality scenario database. The scope of data collection employs two methodologies: knowledge-driven and data-driven approaches. Knowledge-driven methods, based on expert experience, traffic regulations, and ontological tools, are suitable for extracting typical scenarios in structured environments (e.g., highways, intersections). Data-driven approaches further include NDD and accident data. Data quality control is achieved through multi-sensor fusion, deep learning-based trajectory extraction, and ontological knowledge formalization, ensuring scenario structure and interpretability. This chapter's methodology provides a multi-source data integration framework for subsequent scenario generation, enhancing the realism of test scenarios and coverage of critical cases.

## 5 Scenario Generation Methods

All methodologies are based on two types of sources:

knowledge-based and data-driven. According to [47], regardless of the specific method's inclination toward either data source, the connection between them cannot be severed, as it serves to address the limitations inherent in each. This paper will clarify the distinctions between these two approaches.

## 5.1 Traditional Scenario Generation

### 5.1.1 Data pre-processing

A critical pre-processing step is the definition of the ODD, where the environment is broken down into manageable layers and elements (e.g., road, infrastructure, traffic participants) [48]. Continuous parameters within these elements, such as vehicle speed or acceleration, are often discretized into intervals to facilitate combinatorial testing. To prioritize test coverage, a hybrid weighting approach is often employed, combining objective methods like CRITIC (Criteria Importance Through Intercriteria Correlation) for data-driven parameter importance with subjective methods like AHP (Analytic Hierarchy Process) for expert-based judgments. Furthermore, raw trajectory data may undergo significant reconstruction using statistical techniques like wavelet transforms and Nadaraya-Watson kernel regression to mitigate noise and outliers, ensuring data fidelity before scenario extraction [49]. The concept of "representativity" is central; aiming to ensure the statistical distribution of the generated scenario catalog aligns with real traffic conditions, often guided by standardized sampling plans [47], [50].

### 5.1.2 Scenario Extraction

Scenario extraction, or identification, focuses on discerning and formalizing meaningful driving events from the pre-processed data or knowledge base. This stage relies heavily on rule-based systems, logical constructs, and statistical analysis rather than automated feature learning. One powerful method involves partitioning the scenario parameter space into "safe" and "hazardous" zones using quantitative safety metrics like Inverse Time to Collision (ITTC), which effectively extract critical situations from benign ones.

Knowledge-driven approaches utilize expert experience and formalized domain knowledge [51]. For instance, ontology-based models, leveraging Description Logic (DL) and Web Ontology Language (OWL 2), create explicit representations of entities (e.g., vehicles, road parts) and their spatial and semantic relationships, which are extracted from high-resolution maps and perception data [52]. Another model-driven approach, exemplified by TraModeAVTest [53], constructs complex scenarios using Petri nets, where the net structure is derived from the combination relationships of basic traffic regulation scenarios. Statistical anomaly detection is also used to extract "driving anomaly scenarios" by identifying events that exceed predefined thresholds, such as extreme acceleration, which are then manually labeled by experts to attribute causative factors [49]. A systematic literature review highlighted the evolution of this process, suggesting a more formalized seven-step workflow that includes "scenario fusion" to combine scenarios from different sources for better ODD coverage [50].

### 5.1.3 Scenario Derivation

Scenario derivation is the process of generating concrete, executable test cases from the abstracted scenario definitions. This stage employs a variety of traditional techniques to systematically explore the parameter space.

**Combinatorial testing (CT)**

A cornerstone technique used to manage the "combinatorial explosion" of test cases. Improved greedy algorithms, such as the parameter-weighted $\mathrm{ES}(\mathrm{a}, \mathrm{b})$ algorithm, are used to generate an efficient set of test cases that cover all n-way parameter interactions, significantly reducing test volume while maintaining fault detection capability [48]. Following this, statistical sampling methods like Monte Carlo simulation are used for secondary sampling, selecting specific numerical values from the parameter intervals defined by the CT results, often based on their real-world probability distributions. To further enhance efficiency, clustering algorithms like K-means are applied to the sampled values to identify representative hazardous scenarios, using the cluster centers as the final test case parameters [48].

**Rule-Based and Model-Driven Generation**

In more knowledge-intensive frameworks, derivation is guided by formal models and rules. For Petri net-based models, test paths are generated based on coverage criteria (e.g., place-transition coverage), with methods like pairwise combination testing used to handle concurrent structures efficiently. These abstract paths are then instantiated into concrete test cases by assigning values to parameters [53]. Other frameworks use rule-based reasoning systems with Semantic Web Rule Language (SWRL) to deduce candidate-driving behaviors and their priori probabilities based on the extracted scenario characteristics. These are then combined with a posteriori probabilities learned by statistical models like Hidden Markov Models (HMMs) using Bayes's rule to predict behavior, thus deriving the dynamic evolution of a scenario [52].

Table 1. A Detailed Derivation Approaches Comparison

| Approach Category | Core Derivation Technique | Parameter Handling | Explanation |
|---|---|---|---|
| **Statistical / Combinatorial** | Combinatorial Testing (e.g., improved $ES(a,b)$), Monte Carlo Sampling K-means Clustering. | Systematic combination of discretized parameters, followed by random sampling from distributions and clustering to find representative points. | Generating a minimal set of test cases for an AEB system covering various speeds, distances, and weather conditions |
| **Knowledge / Model-driven** | Petri Net Path Generation, Rule-Based Reasoning (e.g., SWRL), Hidden Markov Models (HMMs). | Guided by formal model coverage criteria and logical rules. Parameters are instantiated to satisfy path or rule conditions. | Generating test cases that specifically check for violations of traffic regulations at complex intersections by exploring all valid and invalid behavioral paths |

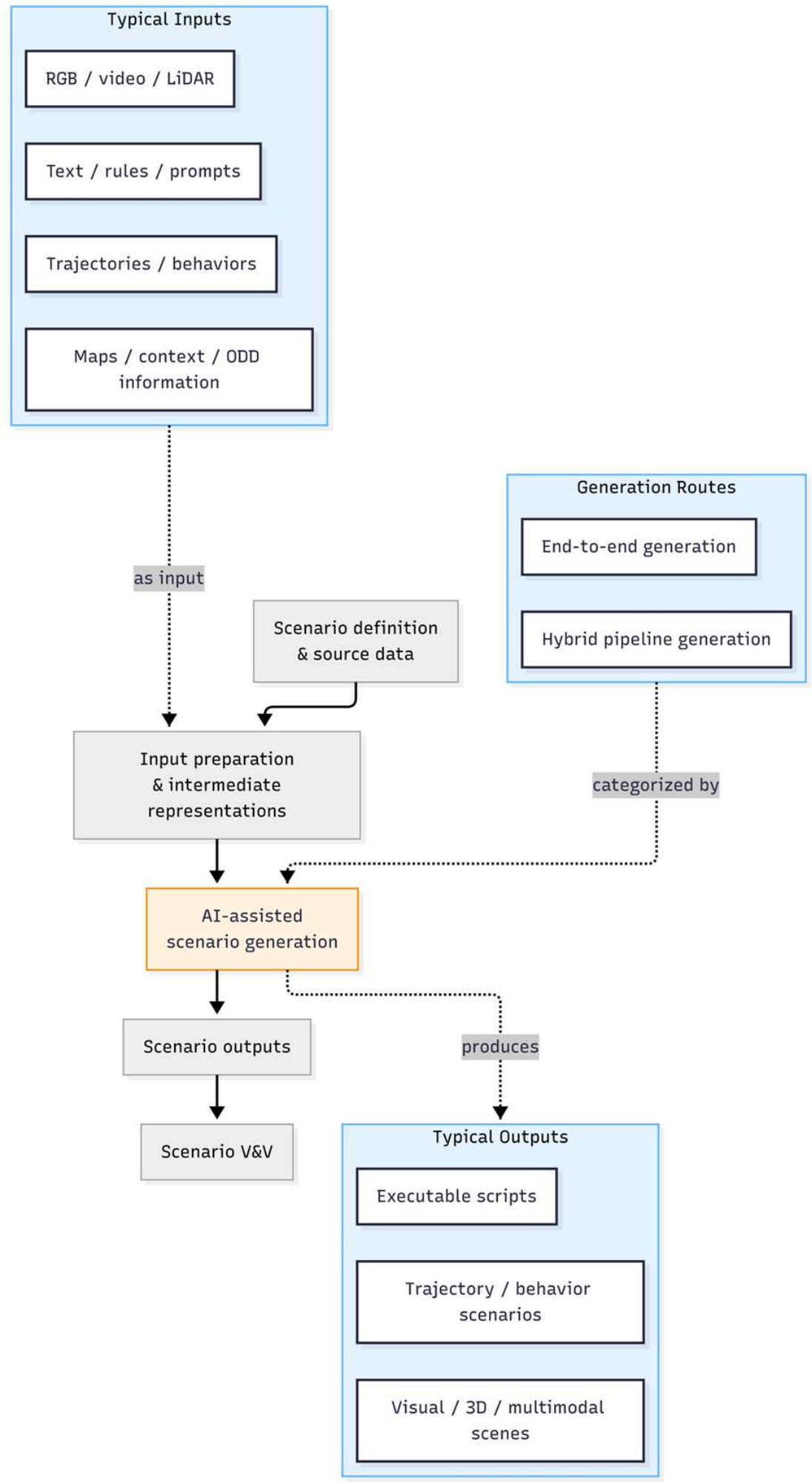

Fig. 4. Workflow of AI-assisted scenario generation for ADS testing

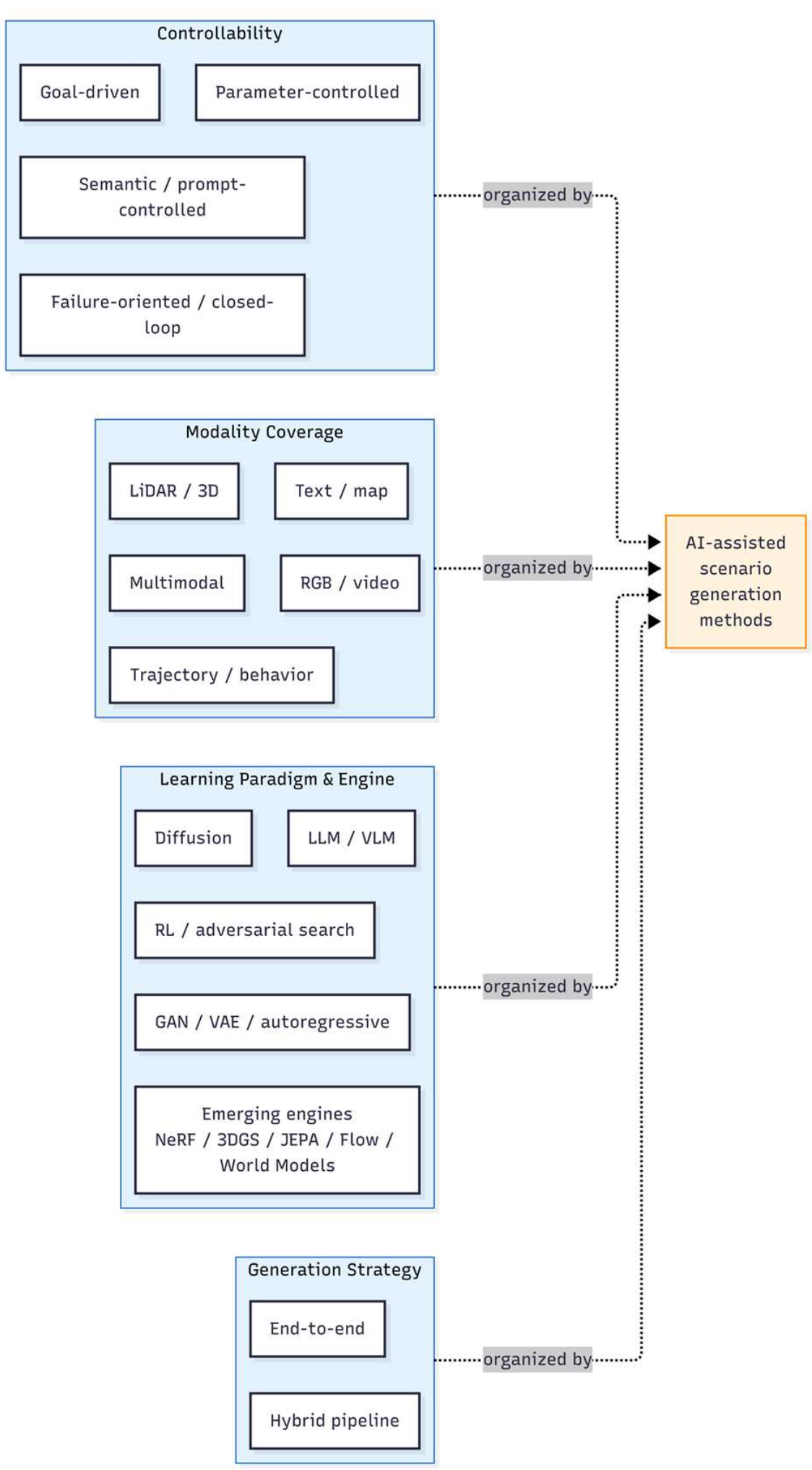

Fig. 5. A multi-axis taxonomy of AI-assisted scenario generation for ADS testing

A comparison of these derivation approaches is summarized in the Table 1 below. This table highlights the fundamental difference in derivation strategy: statistical/combinatorial methods explore the parameter space systematically but agnostically, whereas knowledge-driven methods generate scenarios to satisfy specific logical or behavioral constraints derived from domain expertise.

While traditional approaches such as combinatorial testing and ontology-based reasoning have provided a structured foundation, their limitations are increasingly evident. Rule-based derivation often struggles with scalability when the parameter space expands, and expert-driven ontologies may lag the rapid evolution of traffic regulations or novel driving behaviors. Moreover, statistical sampling methods, although systematic, can generate scenarios that are formally diverse but not necessarily representative of real-world driving distributions. These shortcomings suggest that traditional methods are most effective when used to capture well-understood, regulated environments (e.g., highways), but less suitable for rare or highly interactive urban events.

## 5.2 AI-Assisted Scenario Generation

Fig. 4 summarizes the generic workflow of AI-assisted scenario generation for ADS testing. Compared with traditional derivation pipelines, AI-assisted approaches typically integrate heterogeneous inputs, such as text, maps, trajectories, RGB images, videos, LiDAR, or other structured context, and transform them into executable or perceptual test scenarios through learned generation mechanisms. Depending on the framework, this process may proceed in an end-to-end manner or through hybrid pipelines with intermediate representations. In all cases, generated outputs still require downstream validation and analysis before they can be considered useful for ADS evaluation.

### 5.2.1 Technical Routes and a Multi-Axis Taxonomy

The rapid expansion of AI-assisted scenario generation since 2023 has introduced substantial methodological diversity, making purely architecture-based taxonomies increasingly fragile. For example, frameworks may rely on the same foundation model family while differing substantially in their generation strategy, intermediate representations, modality coverage, and degree of controllability. Such differences directly affect reproducibility, interpretability, and suitability for safety validation, yet they are easily obscured when methods are grouped only by their backbone models.

To address this issue, this section focuses specifically on AI-assisted scenario generation methods rather than attempting to provide a full-spectrum taxonomy of all scenario generation approaches. Traditional methods, including rule-based, ontology-based, and statistical derivation approaches, have been reviewed separately in the preceding subsection. Here, our goal is to organize recent AI-assisted methods in a way that remains compatible with the dominant model-oriented literature while being sufficiently extensible to accommodate emerging paradigms.

Accordingly, Fig. 5 presents an AI-specific multi-axis taxonomy for AI-assisted scenario generation, structured along four complementary dimensions: generation strategy, learning paradigm and engine, modality coverage, and controllability. These dimensions are not intended to replace the familiar model-family view entirely; instead, they provide an interpretive layer that reveals methodological trade-offs more clearly than a single-axis classification.

First, generation strategy distinguishes whether a framework produces scenarios in an end-to-end manner or through a hybrid pipeline with intermediate representations. End-to-end approaches directly map multimodal inputs to executable or perceptual scenarios, which can improve coherence across components but often reduce interpretability when errors occur. Hybrid pipelines, by contrast, introduce intermediate symbolic, geometric, or behavioral representations that support human inspection and modular adaptation, although they may also suffer from error propagation across stages.

Second, learning paradigms and engines characterizes the dominant AI mechanism underlying scenario generation. This dimension includes LLM and VLM-based methods, adversarial and latent-variable generative models, diffusion-based generators, reinforcement-learning-based adversarial search, and emerging predictive or scene-centric AI engines. Within this perspective, architectures such as Neural Radiance Fields (NeRF), 3D Gaussian Splatting (3DGS), flow-based generative models, and JEPA-like predictive frameworks are treated as evolving AI-assisted generation mechanisms rather than isolated top-level categories. This design preserves methodological comparability and is consistent with recent survey efforts that frame foundation models, scene-centric representations, and world-model-like architectures as part of a broader methodological transition in autonomous driving scenario generation and analysis [30], [54], [55].

Third, modality coverage captures the representational scope of a framework. Recent methods increasingly move beyond unimodal trajectory or text generation toward multimodal synthesis involving maps, trajectories, RGB images, videos, LiDAR point clouds, and cross-modal latent representations. This dimension is particularly important for ADS validation because multimodal inconsistency can undermine realism even when individual outputs appear plausible in isolation.

Fourth, controllability describes how scenario generation is driven and constrained. Existing frameworks range from parameter-controlled and template-driven approaches to prompt-driven semantic generation and failure-oriented adversarial generation. This dimension is practically important because controllability determines whether a framework is better suited for routine scenario expansion, targeted safety-critical testing, or closed-loop stress testing.

Under this AI-specific taxonomy, methods are therefore not understood solely by their nominal model family, but by a combination of how they generate scenarios, what modalities they cover, and how users or validation objectives can control them. This perspective also clarifies why recent frameworks often appear hybrid by design: for example, some combine LLM-based semantic parsing with diffusion-based scene synthesis, while others integrate RL with multimodal world generation.

At the same time, model family remains a useful organizational entry point for readers, especially because the current literature from 2023–2025 is still commonly discussed in terms of LLMs, GANs, DMs, and RL. For this reason, the following subsections continue to introduce representative AI-assisted methods according to their predominant generation engines, while interpreting them through the above multi-axis taxonomy. In this way, the review preserves readability and continuity with existing discourse yet avoids over-reliance on a taxonomy that is too tightly coupled to any single generation architecture.

The following subsections therefore discuss representative AI-assisted methods according to their predominant generation engines, while interpreting them through the above multi-axis taxonomy.

### 5.2.2 Transformer-based LLM Approaches

LLMs represent the most prominent models in the field of Artificial Intelligence, with a focus on the text modality and constructed based on the transformer architecture [56]. Most frameworks leverage it, with Chain-of-Thought (CoT) mechanism [57] as a tool to extract useful information from the natural language then utilized in further scenario generation. We adopted the taxonomy in [29] and listed frameworks under the following subtitles.

#### a. Human-machine interface

LLM is responsible for translating natural language into simpler and structured knowledge to enhance the performance of downstream generation work.

b. **Data interpreter**

LLM extracts scenario-relevant information from diverse inputs (including accident reports, domain-specific knowledge, and driving data) and transforms it into natural language representations.

c. **Intermediate format generator**

It decomposes the complex process of generating executable scenario files into multiple stages, producing transitional intermediate representations that integrate both data and knowledge sources.

d. **Standardized format generator**

LLM generates scenario files compliant with standards such as ASAM OpenScenario and AV Unit based on intermediate representations (rather than original scenario sources), utilizing predefined templates or multimodal reasoning.

e. **Executable scenario generator**

LLMs can directly generate executable scenarios from scenario sources in standardized formats (e.g., OpenScenario, SCENIC, SUMO) as well as in user-defined formats.

LeGEND [58] designed a two phase transformation structure, by employing two LLMs to perform the jobs of "Data interpreter" and "Intermediate format generator". This highly integrated design enhanced precision during the transformation. And ADEPT [59] utilized a question-and-answer technique based on the well-known Transformer based model GPT-3 [60] to translate accident report into usable Scenic [61] code for concrete scenario generation. Yet Zhao et al. [62] proposed a framework that leverages the advanced Natural Language Processing (NLP) capabilities of LLM to understand and identify different driving scenarios.

Some of the frameworks adopted advanced adaption techniques greatly improve the effectiveness of LLMs to extract the information, like ChatScene [63] employs GPT-4 alongside Retrieval-Augmented Generation (RAG) to convert textual descriptions of safety-critical matters into the Domain-Specific Language (DSL) for CARLA [64], with a focus on complex urban settings and retrieval database. Further, Tu et al. [65] meticulously designed their prompts using several prompting strategies, in order to enhance the power of LLM in TrafficComposer. Besides, they introduced visual information extractor using YOLOv10 [66], and combining two modalities, resulting in a record that outperforming the best-performing baseline (including TARGET [67]).

Text2Scenario [32] advances this paradigm by introducing a meticulously engineered prompt-based parser that converts free-form textual descriptions—ranging from traffic regulations to accident narratives—into executable OpenScenario files (.xosc) without requiring intermediate structured representations. By decomposing the generation task into semantic parsing, entity extraction, and constraint satisfaction subtasks, the framework achieves high fidelity in translating natural language into machine-executable test cases. The prompt engineering strategy includes few-shot examples and schema-guided generation to reduce hallucination and ensure compliance with OpenScenario syntax. This approach demonstrates that carefully structured prompts can mitigate LLM brittleness yet remain sensitive to linguistic ambiguity and require domain expertise to craft effective prompts for complex or edge-case scenarios.

LLM-driven pipelines remain vulnerable to hallucination, where generated scenarios may be syntactically valid but semantically implausible. Moreover, the lack of standardized prompt engineering practices undermines reproducibility, as minor variations in phrasing can yield substantially different outputs.

**Vision-Language Model (VLM)**

Based on the original LLM architecture, a vision component was embedded to make it able to tell the information holds in an image, creating the Vision Transformer (ViT) [68]. The development of Contrastive Language-Image Pre-training (CLIP) is the milestone of effective zero-shot performance, it makes VLMs more advantageous in the autonomous driving realm. According to its usage, most VLM-based frameworks can be divided into the following types:

- Multimodal context to text
- Image-text matching or enhancing
- Text to image generation

CurricuVLM [69] utilizes Vision-Language Models (VLMs) within an online curriculum learning framework, employing *image-text matching or enhancing* to analyze BEV imagery and task descriptions for safety-critical event identification. Through iterative pattern analysis, GPT-4o generates behavioral diagnostics as *multimodal context translated into plain text* outputs. Complementarily, OmniTester [70] establishes a *text to image generation* pipeline: GPT-4 synthesizes SUMO [71] simulation scripts from user inputs and RAG-augmented external knowledge (e.g., OSM maps), while GPT-4V validates scenarios through visual-code analysis and natural language feedback, demonstrating cross-modal coherence. And DriveGen [72] implements a two-stage traffic scenario generation framework with large model orchestration, integrating map-vehicle asset initialization and trajectory evolution to enhance scenario diversity and realism. Complementarily, the DriveGen-CS module introduces a corner case generation mechanism: leveraging RL from human feedback, it identifies critical scenario parameters (e.g., sudden lane changes, abnormal speed variations) and perturbs baseline trajectories to create safety-critical events. In parallel, Yang et al. [73] developed specialized frameworks for generating critical pedestrian scenarios by integrating trajectory-level conflict modeling with naturalistic behavioral priors, directly addressing the systematic underrepresentation of VRU in automated scenario libraries and achieving validated improvements in VRU interaction coverage. Together, these advances demonstrate greater diversity in edge cases compared to traditional methods.

#### 5.2.3 GAN-based Approaches

GANs [74] have emerged as a necessary technology in advancing ADS testing, particularly in scenario generation. The ability of GANs to approximate complex data distributions has been valuable for enriching scenario diversity. Yet, their training instability and tendency toward mode collapse mean that generated scenarios may not always reflect the full variability of real traffic. In practice, GAN-based methods appear most effective when augmenting existing datasets with stylistic variations (e.g., weather or lighting changes), but less reliable when tasked with producing long-horizon, multi-agent interactions without additional constraints.

The development of GANs in this domain can be categorized by function focus of its application, reflecting their evolving capabilities and integration with other advanced techniques.

**a. Realistic Scene and Image Generation**

These GANs focus on creating visually plausible and diverse static or dynamic driving environments.

Early applications of GANs aimed at synthesizing realistic images and videos to augment training datasets and enhance simulation realism [75]. Early techniques like Pix2PixGAN pioneered paired image-to-image translation, enabling structured transformations (e.g., semantic maps to photos). Simultaneously, CycleGAN introduced unpaired domain adaptation, crucial for tasks like day-to-night conversion without matched datasets, leveraging cycle-consistency losses to preserve content integrity [76]. Recent work like Shi et al. [77] 's dual-branch GAN explicitly decouples depth and RGB information to handle lighting/weather variations. This architecture reduces geometric warping by enforcing depth-awareness, improving structural consistency in autonomous driving test scenarios.

A related but increasingly relevant direction concerns controllable vehicle appearance synthesis rather than full-scene generation. Recent pose-guided methods, such as VehicleGAN, demonstrate that vehicle images can be translated into target poses under both paired and unpaired settings without requiring explicit 3D geometric models [78]. This capability is relevant to AI-assisted scenario construction because it improves the fidelity of viewpoint-varying vehicle assets and can strengthen perception-oriented testing, synthetic dataset augmentation, and cross-camera consistency in traffic scenes. Although such methods do not generate entire scenarios end-to-end, they address a practical bottleneck in scenario realism by improving the controllability and visual consistency of the objects embedded within generated scenes.

**b. Trajectory and Dynamic Interaction Generation**

The shift from static scene generation to dynamic behavior generation is vital for AD testing. Liu et al. [79] utilized Transformer Time-Series Generative Adversarial Networks (TTS-GAN) to generate intersection pre-crash trajectories, addressing the challenge of creating high-risk scenarios with realistic dynamic interactions and long-term temporal dependencies. Liao et al. [80] proposed ITGAN, an Interactive Trajectories Generative Adversarial Network model, which comprehensively considers dynamic interactions between agents and their responses to static road environments.

**c. Safety-Critical and Adversarial Scenario Generation**

Safety-critical scenarios are inherently rare in real-world data, necessitating generative approaches. Stoler et al. [81] introduced Real-world Crash Grounding (RCG), a framework that integrates crash-informed semantics into adversarial perturbation pipelines to generate safety-critical scenarios. Ren et al. [82] leveraged naturalistic driving data to generate high-risk test scenarios at roundabouts, addressing gaps in diversity and realism for extreme conditions. Peng et al. [83] proposed LD-Scene, which combines LLM-guided DMs to controllably generate adversarial safety-critical driving scenarios, highlighting the integration of advanced generative models for precise scenario control.

Although GANs can enrich visual diversity, their outputs often lack long-horizon temporal consistency and may violate physical constraints (e.g., unrealistic accelerations or collision-free assumptions). This limits their applicability for closed-loop ADS validation, where behavioral plausibility is as critical as visual fidelity. In addition, the absence of standardized evaluation metrics for GAN-generated scenarios exacerbates the difficulty of benchmarking their safety relevance.

#### 5.2.4 Diffusion-based Approaches

DMs, fundamentally characterized by a forward process of gradually adding noise to data and a reverse process of denoising to generate samples, have demonstrated exceptional capabilities in producing high-quality and diverse outputs [84], [85], [86]. This generative power has been harnessed to overcome limitations in traditional scenario generation, which often suffers from constrained diversity and realism, particularly for rare and safety-critical events [83], [87]. Yet they exhibit significant potential in joint trajectory prediction and enable more controllable generation compared to conventional deep learning techniques [88], [14], [89] within autonomous driving applications.

**a. High-Fidelity Image and Scene Synthesis**

Early advancements in DMs, such as Denoising Diffusion Probabilistic Models (DDPMs), established a robust framework for generating high-quality samples [84]. These models operate directly in pixel space, offering state-of-the-art synthesis results [85]. Latent Diffusion Models (LDMs) emerged as a key innovation, decomposing the image formation process into sequential denoising autoencoders within a compressed latent space. This approach significantly reduces computational costs associated with high-resolution image synthesis while maintaining high fidelity and enabling control over the generation process without extensive retraining. For instance, models like Imagen leverage large transformer language models for deep language understanding combined with DMs for photorealistic image generation, proving surprisingly effective for encoding text for image synthesis [90]. Further refinements have enabled context-aware image generation, where models learn from visual examples presented in context, improving generation quality and fidelity [91]. This is crucial for synthesizing complex scenes that are consistent with given visual cues. The development of controllable arbitrary viewpoint camera data generation, such as **ArbiViewGen**, addresses the lack of ground-truth data in extrapolated views by introducing feature-aware and geometry-aware components within a diffusion-based framework [92]. DriveGEN [93] provided another approach of scenario image generation by requiring no training process and relying on relatively coarse self-attention features. This capability is vital for creating diverse visual perspectives for ADS training and testing.

A notable industrial contribution to this domain is GAIA-1 [13], a generative world model developed by Wayve that synthesizes photorealistic, multimodal driving videos from text prompts, image inputs, or sensor data. By combining video, text, and action inputs in an autoregressive transformer architecture, GAIA-1 demonstrates the ability to generate spatially and temporally consistent driving scenarios, including diverse weather conditions, road topologies, and dynamic agent interactions. Its capacity for controllable generation—where users can specify high-level scene attributes via natural language—positions it as a pioneering framework for simulator-augmented ADS validation. However, the model's reliance on proprietary datasets and undisclosed architectural details limits independent reproducibility and raises questions about generalizability beyond the training distribution.

**b. Controllable Scenario Generation**

A primary breakthrough has been the development of DMs for controllable scenario generation, allowing for the synthesis of specific and complex driving environments. DiffRoad [94], for example, is a novel DM designed to produce controllable and high-fidelity 3D road scenarios, synthesizing road layouts with realistic elements and varying traffic conditions. This facilitates comprehensive testing and validation of autonomous vehicles by enabling the creation of authentic and varied scenarios that are challenging to acquire from real-world data alone. DiffScene [95] designed several adversarial optimization objectives to guide the diffusion generation under predefined adversarial budgets, making it more transferable to different AV algorithms. Similarly, Scenario Diffusion [96] introduces a diffusion-based architecture that enables controllable scenario generation by combining latent diffusion, object detection, and trajectory regression to simultaneously generate distributions of synthetic agent poses, orientations, and trajectories. This approach provides additional control over generated scenarios, making it invaluable for validating the safety of autonomous vehicles. Scenario Dreamer [97] offers a fully data-driven generative simulator that generates both the initial traffic scene, including lane graphs and agent bounding boxes, and closed-loop agent behaviors, overcoming the inefficiencies of rasterized image representations. Versatile Behavior Diffusion (VBD) [98] frames imitation learning as diffusion-based generative modeling, enabling the generation of realistic and controllable agent behaviors in traffic simulations.

Diffusion-based approaches have shown strong potential in generating high-fidelity and controllable scenarios. Nevertheless, their reliance on large-scale training data and high computational budgets limits accessibility, and the interpretability of generated output remains an open question. For safety validation, DMs may need to be paired with rule-based filters or physics-informed constraints to ensure that rare but safety-critical cases are not only visually plausible but also behaviorally consistent with traffic dynamics. Without physics-informed constraints or post-generation filtering, diffusion-based scenarios risk being visually plausible but dynamically inconsistent, undermining their value for safety-critical validation.

#### 5.2.5 Auxiliary Techniques

In the field of scenario generation for ADS, AI-based auxiliary methods—particularly Recurrent Neural Networks (RNNs) and Reinforcement Learning (RL)—have made significant strides over the past decade. These advancements enhance the realism, diversity, and scalability of virtual simulations, enabling effective testing and validation. Traditional scene generation methods rely on rule-based systems, knowledge-driven models, and data-driven synthesis, often exhibiting limitations in diversity and in terms of realism for safety-critical scenarios. Therefore, integrating AI technologies such as RNNs and RL with other advanced techniques—including GANs, VAEs, and LLMs—is crucial for constructing robust and scalable simulation environments [30].

RL-based adversarial generation is extremely sensitive to reward shaping; poorly designed reward functions may be overfit to artificial failure modes rather than uncovering realistic vulnerabilities. Furthermore, policies trained in simulation often exhibit limited transferability to real-world distributions, raising questions about their external validity.

**a. Recurrent Neural Network**

RNNs excel at modeling temporal dependencies in driving scenarios, enabling realistic trajectory prediction and sequential data synthesis.

**b. Trajectory Synthesis with Hybrid Architectures.**

RNNs (e.g., LSTM/GRU) integrate with GANs to generate variable-length vehicle trajectories. Demetriou et al. work combines conditional GANs with RNNs to synthesize multi-agent interactions, ensuring temporal consistency in traffic flow [99]. Models like these encode historical trajectory data and decode future paths, improving scenario diversity versus rule-based methods.

**c. Enhancing LLM-Based Scene Description**.

RNNs process LLM-generated textual scene descriptions (e.g., "lane change during heavy rain") into structured temporal sequences. Peng et al.'s LD-Scene uses RNNs to refine LLM outputs into actionable parameters for DMs [83]. RNNs align semantic LLM prompts with low-level sensor data (e.g., LiDAR point clouds), enabling coherent scenario augmentation [100].

**d. Reinforcement Learning (RL)**

RL optimizes scenario generation through trial-and-error, prioritizing adversarial or high-risk conditions.

Zhang et al. [101] deploy RL agents to perturb initial scenarios (e.g., weather parameters, agent behaviors), maximizing ADS failure rates. Reward functions incorporate risk metrics like Time-to-Collision (TTC), generating more corner cases than random sampling.

Xu et al. [102] use RL to optimize LLM prompts for scenario diversity. Agents receive rewards for generating rare events (e.g., pedestrian jaywalking) validated in simulation. RL refines DMs by rewarding photorealistic hazard scenarios (e.g., blurred visibility, dynamic obstacles). Yang et al.'s neural rendering approach shows a 41% gain in scene realism [103].

### 5.2.6 Summary

In summary, AI-assisted methods substantially advance scenario diversity and multimodal integration, yet their adoption is constrained by interpretability gaps, reproducibility challenges, and heavy dependence on proprietary resources. These limitations highlight the necessity of hybrid frameworks that combine the transparency and regulatory alignment of traditional approaches with the generative flexibility of AI, while embedding safeguards to mitigate hallucination, instability, and dataset bias.

# 6 Evaluation and Comparison

In this section, we will introduce evaluation metrics and conduct a comparative analysis of various generative frameworks.

## 6.1 Metrics

In this review, three comparative evaluation metrics are employed to ensure comparability across scenario generation methods. These metrics are designed to reflect scholarly impact, reproducibility and ethical responsibility, and the intrinsic quality of generated scenarios, respectively.

At the current stage, these metrics should be understood as review-oriented comparative instruments rather than empirically validated predictors of downstream ADS safety performance. These metrics characterize upstream properties of scenario-generation frameworks—such as scholarly influence, reproducibility, and ODD coverage—rather than measuring the downstream behavioral safety performance of an ADS. In this review, downstream safety indicators refer to planner- or decision-level behavioral signals observed during closed-loop execution (e.g., collision discovery rate, policy intervention frequency, minimum time-to-collision, and rule-violation count). Their main purpose is to support transparent cross-framework comparison in rapidly evolving literature, while leaving benchmark-oriented validation to future work. Each metric is defined by explicit parameters and weighting schemes to maintain transparency and credibility.

### 6.1.1 Academic Influence Index (AII)

The Academic Influence Index (AII) is formulated to account for the prevalence of preprints in this domain, where traditional venue-based indicators are insufficient. It is defined as:

$$AII = \frac{1}{3} \times (\mathrm{C}_{norm} + \mathrm{C}_{\mathrm{y1}} + \mathrm{H}_{\mathrm{mean}}). \quad (1)$$

To ensure an objective and impartial evaluation of scholarly impact, the weighting coefficients for the AII are assigned an **equal distribution** (1/3 for each component). This balanced configuration recognizes that in the rapidly evolving domain of AI-assisted ADS testing, academic influence is inherently multi-dimensional. Citation counts ($C_{norm}$) represent established recognition, while the rate of early dissemination ($C_{y1}$) is crucial for assessing the impact of preprints— the first-year citation share, it is also a dominant medium in current AI research. Furthermore, incorporating the authorial track record ($H_{mean}$) provides a proxy for the reliability and sustained contribution of the research team. By employing equal weights, the AII provides a neutralized and comprehensive assessment that avoids over-reliance on any single metric, thereby ensuring a fair comparison between legacy publications and emerging high-impact preprints.

### 6.1.2 Resource Accessibility Score (RAS)

To quantify the reproducibility and practical utility of each framework, we define the Resource Accessibility Score (RAS). RAS evaluates the availability of essential artifacts required for independent verification, ranging from source code to environmental configurations. The total score is a weighted aggregation of five key dimensions: Code Availability ($S_{code}$), Minimal Reproducible Pipeline ($S_{pipe}$), Environment Transparency ($S_{env}$), Model and Asset Accessibility ($S_{asset}$), and Documentation Quality ($S_{doc}$).

$$RAS = \min\left(100, 100 \times \sum_{i=1}^{5} w_i \cdot s_i + \mathrm{B}_{dataset}\right) \quad (2)$$

To ensure maximum transparency and reproducibility of our evaluation, each dimension is assessed based on standardized scoring rubric with predefined criteria. The specific scoring logic and weights ($w_i$) and the formalized calculation procedure are provided in Appendix C to create a reproducible basis for the comparative evaluation.

To improve interpretability, RAS is mapped to a five-star scale based on the final score:

Table 2. Star-Based Representation of Resource Accessibility Scores

| Score Range | Star Level |
|---|---|
| **≥90** | ***** Excellent |
| **75–89** | **** Good |
| **60-74** | *** Moderate |
| **40-59** | ** Limited |
| **<40** | * Poor |

This star-based representation provides an intuitive summary of resource accessibility while preserving quantitative rigor.

### 6.1.3 ODD Coverage Score (OCS)

The ODD Coverage Score (OCS) quantifies the breadth of operational conditions addressed by a scenario generation framework. Following the rubric in Appendix D, OCS evaluates six dimensions: Road Type Coverage, Vulnerable Road User (VRU) Presence, Topological Complexity, Interaction Complexity, Scenario Controllability, and Regional Traffic Regulation Diversity. Each dimension is scored on a five-level scale (0.10–1.00) with **equal weights** (16.7% each) and explicit anchors to minimize subjectivity. The six dimensions assess orthogonal aspects of scenario coverage: environmental diversity, agent complexity, user flexibility, and regulatory context.

The overall OCS is computed as:

$$OCS = 100 \times \left( \frac{1}{6} \sum_{k=1}^{6} a_k \right) \quad (3)$$

Scoring Rules based on a tiered evidence strategy. Specific scoring criteria are described in Appendix D.

To ensure interpretability and consistency with RAS, the percentage score is mapped to the five-star scale (Table 2).

### 6.1.4 Metrics Usage

This subsection demonstrates the application of the proposed metric suite (AII, RAS, OCS) by evaluating six representative frameworks spanning diverse technical routes and performance profiles: **GAIA-1** (DM, industrial benchmark), **TrafficGEN** (VAE/Autoregressive, trajectory-focused), **Text2Scenario** (LLM, language-driven), **SEAL** (Reinforcement Learning, adversarial), **Genesis** (DM, multimodal consistency), and **TrafficComposer** (LLM+VLM, hybrid generation). These frameworks cover five distinct generative paradigms and exhibit substantial variation across all three metric dimensions, illustrating the complementary insights provided by AII, RAS, and OCS. All metric inputs are collected from published papers, public repositories, Google Scholar profiles (accessed September 2025), and project documentation. Where papers present qualitative claims (e.g., "supports diverse weather conditions"), we translate them conservatively into the OCS rubric and cite specific figures, tables, or code repositories as evidence.

a. **Academic Influence Index**

The Academic Influence Index is computed as follows: $AII = \frac{1}{3} \times (C_{norm} + C_{y1} + H_{mean})$, where $C_{norm}$ represents citation count normalized against the maximum observed value (**GAIA-1**: 359 citations), $C_{y1}$ denotes the first-year citation share, and $H_{mean}$ captures the mean h-index of primary authors normalized by the highest team value (**Genesis**: mean 71). This equal-weight formulation ensures that no single metric dominates the assessment, providing a balanced evaluation of scholarly impact across established recognition, recent dissemination velocity, and authorial credibility. All bibliometric data were collected from Google Scholar in September 2025.

**GAIA-1** leads the evaluated set with 359 citations accumulated since its September 2023 release. Estimating first-year citations at approximately 150 (42% of total) and principal investigator h-index at 35, the calculation yields $C_{norm}$ = 1.0, $C_{y1}$ ≈ 0.42, $H_{mean}$ ≈ 0.49, producing AII = 63.7%. This framework represents a mature industrial contribution from Wayve, combining established citation momentum with high-profile authorship.

**TrafficGEN**, presented at ICRA 2023, has attracted 147 citations with an estimated first-year share of 48%. The lead author's h-index of approximately 20 produces $C_{norm}$ ≈ 0.41, $C_{y1}$ ≈ 0.48, $H_{mea}$ ≈ 0.28, yielding AII = 39.0%. The framework demonstrates strong uptake within the trajectory generation subcommunity, reflected in its citation velocity and reproducibility-focused implementation.

**Text2Scenario** achieved eleven citations within six months of its March 2025 publication (100% first year). With estimated mean author h-index of 25, the calculated values $C_{norm}$ ≈ 0.031, $C_{y1}$ = 1.0, $H_{mean}$ ≈ 0.35 produce AII = 46.0%. The rapid uptake reflects methodological interest in prompt-based scenario parsing, where natural language interfaces address a recognized gap in testing workflows. This case illustrates how recent work can achieve moderate influence through timely problem selection rather than author prestige alone.

**SEAL** accumulated seven citations by September 2025 (all first-year). An estimated author h-index of 22 yields $C_{norm}$ ≈ 0.019, $C_{y1}$ = 1.0, $H_{mean}$ ≈ 0.31, producing AII = 44.3%. The framework occupies the mid-tier typical of specialized RL contributions: solid methodological interest within adversarial testing communities but limited broader diffusion due to narrow applicability and reproduction barriers.

**Genesis** received only 1 citation as of evaluation but achieves AII = 33.4% almost entirely through senior-author standing. The author team—Wenyu Liu (h=89), Xinggang Wang (h=85), Guang Chen (h=39)—contributes via the h-index term ($H_{mean}$ = 1.0), whereas citation components remain near zero ($C_{norm}$ ≈ 0.003, $C_{y1}$ = 0). This profile is characteristic of very recent publications by established research groups, where anticipated impact precedes empirical validation. The AII will shift substantially as citation counts stabilize over the coming year.

**TrafficComposer** shows zero stable citations at evaluation time. The author team (Tianyi Zhang: h=26, Zhi Tu: h=8) produces mean h-index 17, yielding $H_{mean}$ ≈ 0.24 and AII = 8.0% purely from the h-index component ($C_{norm}$ = 0, $C_{y1}$

= 0). This "cold start" affects emerging research teams irrespective of technical merit, motivating the inclusion of RAS and OCS as complementary axes that assess reproducibility and scenario coverage independently of citation networks.

The distribution spans three tiers: works with mature citation histories (**GAIA-1**: 63.7%), recent contributions by mid-career authors (**Text2Scenario**: 46.0%, **SEAL**: 44.3%, **TrafficGEN**: 39.0%, **Genesis**: 33.4%), and emerging team outputs (**TrafficComposer**: 8.0%). **Genesis** and **TrafficComposer**—both released mid-2025 within one month—differ by twenty-five percentage points solely through author h-index disparity (71 vs. 17), illustrating AII's sensitivity to academic pedigree. This variation provides empirical justification for integrating RAS and OCS, which measure reproducibility and operational coverage orthogonally to scholarly metrics, thereby preventing citation-based bias from obscuring technically sound but less-established contributions.

b. **Resource Accessibility Score**

RAS aggregates five weighted components—Code Availability (30%), Minimal Reproducible Pipeline (25%), Environment/Build Transparency (20%), Model/Asset Accessibility (15%), and Documentation Quality (10%). Detailed scoring criteria for each dimension are provided in Appendix C. Each component is scored on {1.0, 0.5, 0} based on the completeness criteria defined therein. A dataset bonus $B \in \{0, +2, +5\}$ rewards use of public benchmarks. Final scores map to a five-tier star system: ≥90% earns 5*, 75-89% earns 4*, 60-74% earns 3*, 40-59% earns 2*, and <40% earns 1*.

**TrafficGEN** provides complete code via GitHub, including pre-trained VAE weights, Waymo Open Dataset loaders, and one-command demonstration scripts. Environment specifications are documented through requirements.txt, and model checkpoints are publicly hosted. We assign Code (1.0), Pipeline (1.0), Environment (1.0), Model/Assets (1.0), Documentation (1.0), and apply the "+5" dataset bonus for Waymo integration, yielding RAS = 105%, capped at 100% (5*). This framework sets the reproducibility benchmark among evaluated methods.

**TrafficComposer** offers an end-to-end runnable repository with code, pipeline execution scripts, and pointers to public pretrained models (YOLOv10, CLRNet). However, containerization (Docker/Singularity) and dependency lockfiles are absent, introducing minor environment variability. Scoring Code (1.0), Pipeline (1.0), Environment (0.5), Model/Assets (1.0), Documentation (1.0) produces RAS = 90% (5*).

**GAIA-1**, developed by Wayve, is published as a technical report with research documentation and demonstration videos available on the project page. No source code, executable pipelines, or model weights are released. We score Code (0), Pipeline (0), Environment (0), Model/Assets (0), Documentation (1.0), with no dataset bonus, producing RAS = 10% (1*). The absence of implementation artifacts reflects commercial intellectual property protection typical of industry-backed research, where proprietary technology precludes independent verification.

**Text2Scenario** distributes prompt templates, few-shot examples, and text-to-OpenScenario conversion scripts via GitHub. The pipeline depends on external GPT-4 API access, and environment setup for OpenScenario validators is left to users. Scores are assigned as Code (1.0), Pipeline (0.5), Environment (0.5), Model/Assets (1.0), and Documentation (1.0), with no dataset bonus for zero-shot operation, yielding RAS = 77.5% (4*). The reliance on proprietary APIs introduces access barriers but does not fundamentally compromise reproducibility.

**SEAL** includes partial training loop code and adversarial scenario generation scripts, but lacks pre-trained policies, simulator integration documentation, and hyperparameter sweep details. Manual configuration of CARLA or SUMO environments is required, and no containerization is provided. We assign Code (0.5), Pipeline (0.5), Environment (0), Model/Assets (0.5), Documentation (0.5), and no dataset bonus, producing RAS = 40% (2*). These limitations reflect widespread RL reproducibility challenges: environment sensitivity, policy non-transferability, and simulator dependency.

**Genesis** hosts a public GitHub repository with README and licensing but announced models and runnable pipelines as "coming soon" at evaluation time. We score Code (0.5), Pipeline (0), Environment (0), Model/Assets (0), Documentation (0.5), and apply +5 for nuScenes public dataset usage, yielding RAS = 25% (1*). The incomplete release diminishes practical utility despite strong author credentials (AII = 33.4%).

RAS scores span the full 1*-to-5* spectrum. TrafficGEN (100%) and TrafficComposer (90%) demonstrate that well-resourced academic teams can achieve near-complete reproducibility when incentivized. Text2Scenario (77.5%) falls into the 4* tier, where external API dependencies introduce access barriers but do not fundamentally compromise reproducibility. SEAL (40%) and Genesis (25%) represent the "promise-delivery gap": high academic standing or methodological innovation does not ensure resource accessibility. GAIA-1 (10%) occupies the lowest tier, where commercial confidentiality prevents code release entirely. This stratification validates RAS's utility in distinguishing frameworks suitable for independent adoption from those requiring institutional partnerships or remaining fully proprietary.

c. **ODD Coverage Score**

The ODD Coverage Score evaluates six dimensions: Road Type Coverage, VRU Presence, Topological Complexity, Interaction Complexity, Scenario Controllability, and Regional Traffic Regulation Diversity. Each dimension is scored on [0.1, 0.25, 0.5, 0.75, 1.0] according to Appendix D. The overall score is the weighted average across six dimensions (equal weights: 16.7% each). To ensure interpretability and consistency with RAS, the percentage score is mapped to the five-star scale (Table 2).

**GAIA-1** video demonstrations showcase diverse scenarios including urban streets, highways, rain/fog conditions, day/night cycles, and multi-agent interactions (vehicles + pedestrians). Text/image/action conditioning enables fine-grained control over scene composition. We assign Road Type (0.75), Interaction Complexity (0.75), Scenario Controllability (1.0), VRU Presence (0.75), and Topology (0.75), producing strong coverage across environmental and interaction dimensions. However, training data originates exclusively from Wayve's UK operations (left-hand traffic), limiting validated applicability to other regulatory contexts; we assign Regional Diversity (0.25). Overall OCS = 71% (3*). The world-model architecture inherently supports broad ODD coverage, constrained primarily by training data distribution.

**Text2Scenario** natural language interface supports arbitrary scenarios expressible in text. Paper examples include accident reports (rear-end collisions, lane intrusions), traffic violations (red-light running), and adverse weather narratives. We assign Road Type (0.75), Interaction Complexity (0.75), Scenario Controllability (1.0), VRU (0.75), and Topology (0.75). LLM-based architecture theoretically enables region-specific scenario generation, yet the paper does not demonstrate explicit adaptation to diverse traffic regulations (e.g., left-hand versus right-hand driving, roundabout priority rules); we assign Regional Diversity (0.25) based on verifiable evidence. It is overall OCS = 71% (3*). Coverage is limited primarily by prompt engineering creativity and LLM hallucination risks.

**TrafficComposer** benchmarks include highway and urban roads across nine cities in five countries, as stated in the paper. Natural language and reference images are interpreted into executable scripts. We scored Road Type (0.50), Interaction Complexity (0.75), Scenario Controllability (0.75), VRU (0.50), and Topology (0.75). While the paper claims multi-national coverage, specific country names and regulatory adaptations are not enumerated; applying the verifiable-evidence principle, we assign Regional Diversity (0.25). The overall OCS = 59% (2*). The framework balances breadth with simulator-oriented constraints.

**SEAL** focuses on adversarial corner-case discovery via RL, deliberately prioritizing safety-critical events over ODD breadth. Paper experiments concentrate on highway lane-change conflicts and intersection near-miss scenarios within CARLA simulator. We assign Road Type (0.50), Interaction Complexity (0.75), Scenario Controllability (0.50), VRU (0.25), and Topology (0.50). The framework employs default CARLA traffic rules without demonstrating region-specific configurations; we assign Regional Diversity (0.10). The overall OCS = 40% (2*). The lower score reflects narrow, deep specialization deliberate trade-off for safety validation.

**Genesis** paper and project page show urban scenes with straight city streets, multi-arm intersections, and T-junctions (Fig. 7, Fig. 12-15). Editable captions modify weather and time (Fig. 9). Multiple vehicles and pedestrians coexist with cross-modal LiDAR consistency (§4.5, Fig. 12-15). We scored Road Type (0.25), Interaction Complexity (0.25), Scenario Controllability (0.75), VRU (0.25), and Topology (0.50). Genesis uses the nuScenes dataset, which comprises data from Boston, USA (right-hand traffic), and Singapore (left-hand traffic), making it the only evaluated framework with validated coverage of both driving directions; we assign Regional Diversity (0.75). The overall OCS = 44% (2*).

**TrafficGEN** generates realistic multi-agent trajectories but focuses narrowly on highway cruising and lane-change maneuvers. The VAE latent space offers limited controllability, and VRU scenarios are sparse. We assign Road Type (0.25), Interaction Complexity (0.50), Scenario Controllability (0.25), VRU (0.10), and Topology (0.25). Training data derives exclusively from Waymo Open Dataset collected in the western United States (right-hand traffic); we assign Regional Diversity (0.25). The overall OCS = 27% (1*). The low score reflects coverage limitations of trajectory-only generation without full scene context.

OCS scores under the six-dimension rubric stratify into three tiers: moderate-coverage frameworks (**GAIA-1**, and **Text2Scenario**: 71%), lower-moderate frameworks (**TrafficComposer**: 59%, **Genesis**: 44%, **SEAL**: 40%), and narrow-focus specialists (**TrafficGEN**: 27%). **GAIA-1** and **Text2Scenario** achieve identical overall scores (71%) despite differing profiles: **GAIA-1**'s world-model architecture produces strong environmental diversity but is constrained by UK-only training data (Regional Diversity: 0.25), while **Text2Scenario**'s language interface offers broad controllability yet lacks demonstrated region-specific adaptation (Regional Diversity: 0.25). **Genesis**, though scoring lowest in most dimensions, achieves the highest Regional Diversity (0.75) by virtue of nuScenes' Boston-Singapore coverage—the sole evaluated framework validated across left-hand and right-hand traffic contexts. **TrafficComposer**'s decline from moderate coverage (62% under five dimensions) to 59% reflects unverified multinational claims, illustrating the importance of evidence-based regional assessment. This divergence underscores that ODD coverage is multi-faceted: environmental breadth, interaction complexity, and regulatory diversity constitute orthogonal dimensions requiring independent evaluation.

Table 3. Six-framework Evaluation Results

| Framework | AII | RAS | OCS | Regional Diversity | Key Characteristics |
|---|---|---|---|---|---|
| **GAIA-1** | 63.7% | 10% (1*) | 71% (3*) | 0.25 | Commercial; minimal resource accessibility |
| **Text2Scenario** | 46.0% | 77.5% (4*) | 71% (3*) | 0.25 | LLM; high controllability; regional undemonstrated |
| **TrafficComposer** | 8.0% | 90% (5*) | 59% (2*) | 0.25 | Multi-national unverified; high reproducibility |
| **Genesis** | 33.4% | 25% (1*) | 44% (2*) | 0.75 | Best regional diversity (Boston + Singapore) |
| **SEAL** | 44.3% | 40% (2*) | 40% (2*) | 0.10 | Adversarial RL; CARLA default; narrow ODD |
| **TrafficGEN** | 39.0% | 100% (5*) | 27% (1*) | 0.25 | Perfect reproducibility; US Waymo; trajectory-only |

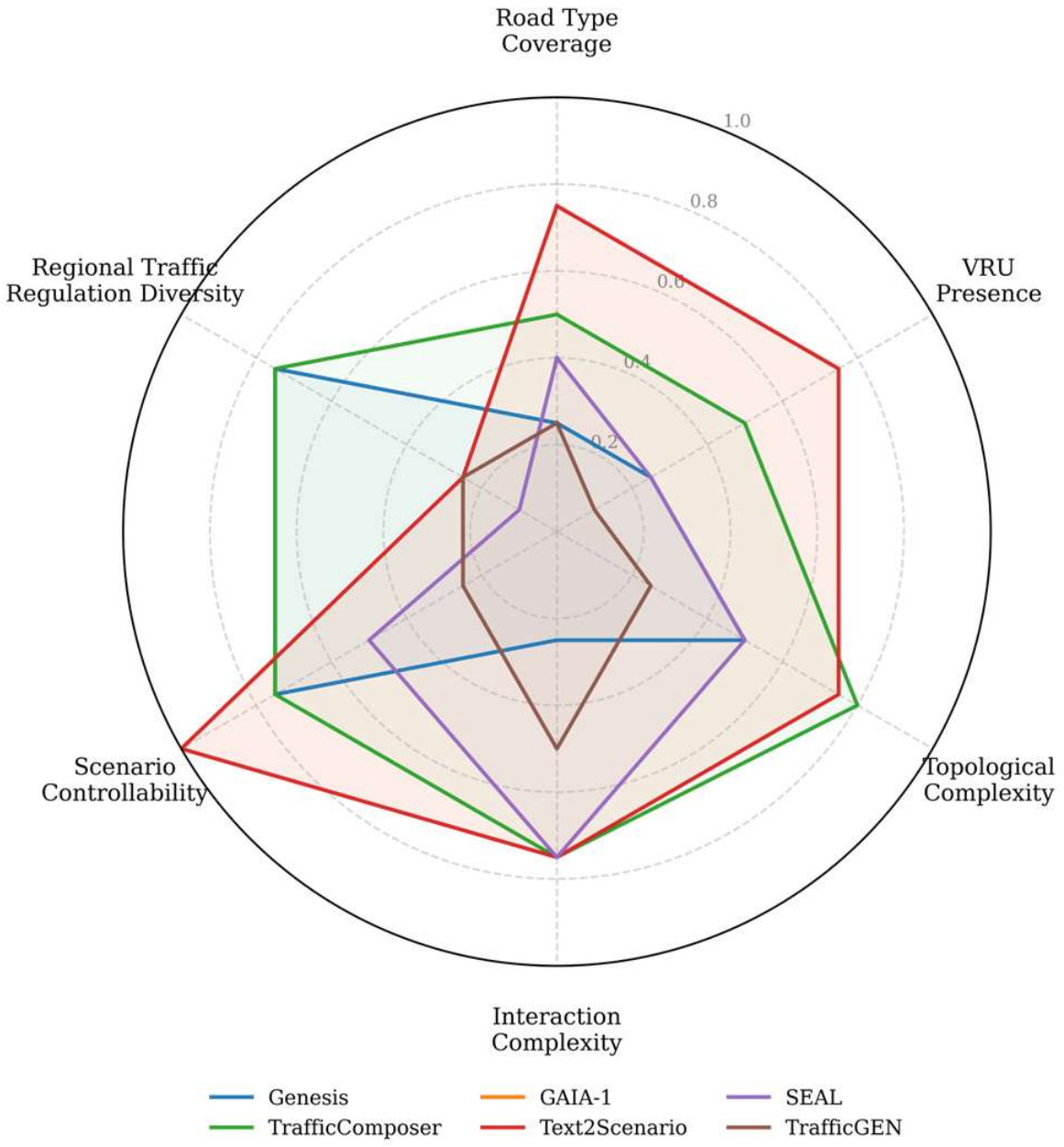

Fig. 6 OCS sub-dimension capability assessment of different frameworks

d. **Three-Metric Synthesis**

Table 3 consolidates the six-framework evaluation, revealing complementary insights across metrics. **GAIA-1** demonstrates the strongest overall performance (AII 63.7%, RAS 10%, OCS 71%), achieving high scores across academic impact, reproducibility, and scenario coverage. And the bar chart (Fig. 7) displays the evaluation results of six frameworks across three key metrics: Academic Influence Index (AII), Resource Accessibility Score (RAS), and ODD Coverage Score (OCS). The radar chart (Fig. 6) illustrates the performance distribution across six specific dimensions within six different frameworks: Road Type Coverage, VRU Presence, Topological Complexity, Interaction Complexity, Scenario Controllability, and Regional Traffic Regulation Diversity.

**TrafficGEN** exhibits perfect reproducibility (RAS 100%) and moderate influence (AII 39.0%) yet achieves minimal ODD coverage (OCS 27%), distinguishing technical excellence from scenario versatility. **Text2Scenario** demonstrates that moderate academic impact (AII 46.0%) can coexist with strong ODD coverage (OCS 71%), provided the methodology prioritizes user-facing flexibility, though regional validation remains limited.

**Genesis** and **TrafficComposer**—both mid-2025 releases—occupy opposite RAS extremes (25% vs. 90%), illustrating that publication timing alone does not dictate resource accessibility; team practices and incentive structures dominate. **SEAL**'s uniformly moderate scores (AII 44.3%, RAS 40%, OCS 40%) reflect the RL paradigm's inherent trade-offs: adversarial specialization at the expense of broad coverage, and simulator dependency hindering reproducibility.

These results substantiate the proposed metric suite's core thesis: no single dimension suffices to rank scenario generation frameworks. AII rewards scholarly visibility but overlooks emerging teams (**TrafficComposer**) and penalizes recent releases (**Genesis**). RAS exposes reproducibility gaps (**Genesis** 25%) even among high-AII works. OCS identifies coverage limitations (**TrafficGEN** 27%) invisible to citation counts or code availability. By triangulating these three perspectives, the framework provides multidimensional assessment aligned with diverse stakeholder priorities: researchers prioritizing influential methodologies (AII), engineers seeking deployable tools (RAS), and validation teams requiring comprehensive test

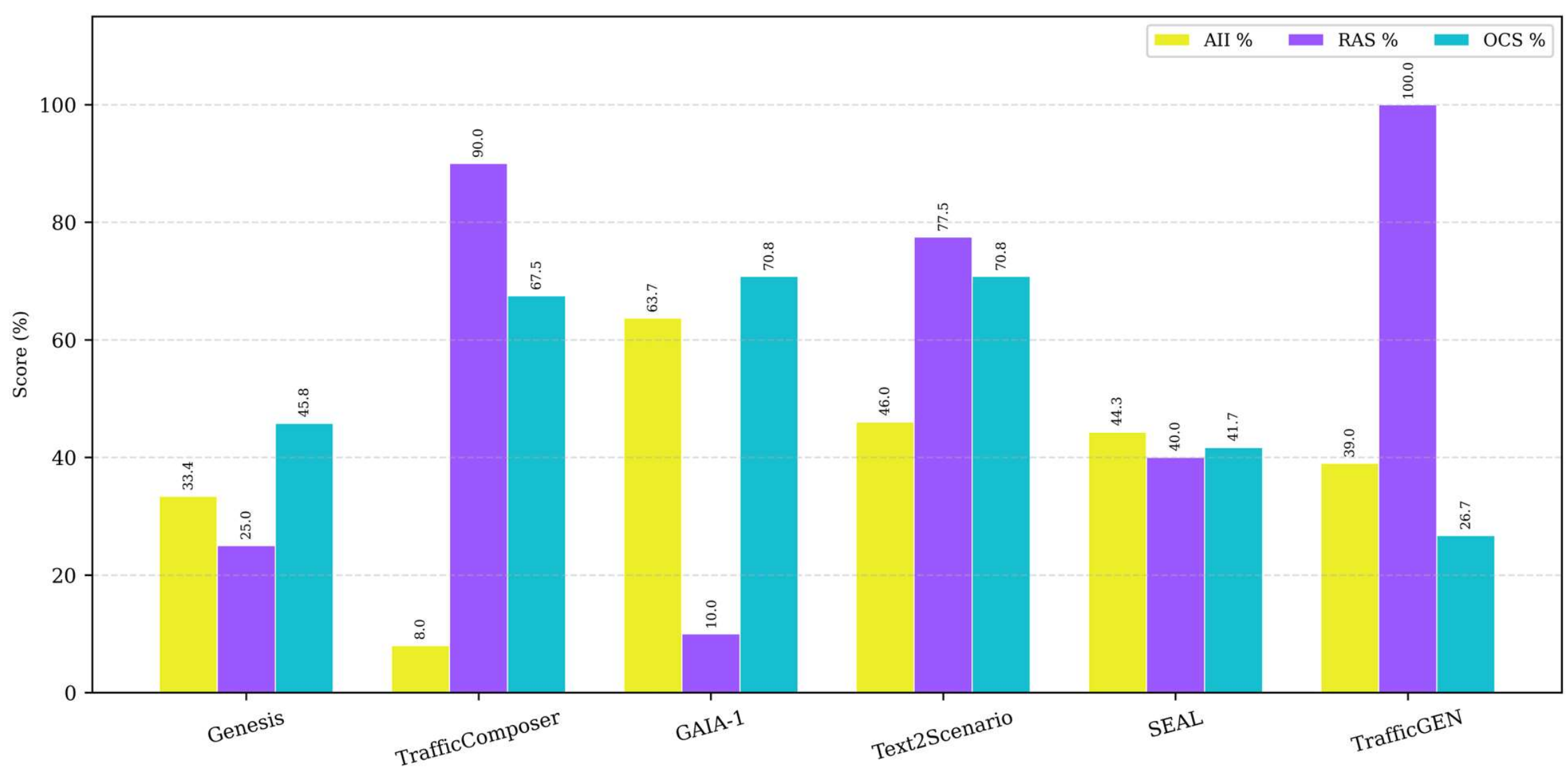

Fig. 7 Quantitative comparison of scenario generation performance

suites (OCS).

These case-based applications are intended to illustrate how the metric suite can structure comparative interpretation across heterogeneous frameworks. They are not, at this stage, substitutes for controlled benchmarking against standardized downstream safety outcomes.

## 6.2 Comparative Table

### 6.2.1 Input and Output Modalities

a. **Visual**

Refers to static image data, commonly used to provide background information on scenes, road environments, traffic signs, static obstacles, etc. A continuous sequence of images capable of capturing dynamic information such as the movement states of vehicles, pedestrians, and cyclists; changes in traffic flow; and environmental dynamics (e.g., weather, lighting variations) is also included. Video modality offers unique advantages in generating dynamic scenes and evaluating autonomous driving systems' responses to time-series events.

b. **3D Sensing**

Primarily sourced from LiDAR sensors, Point-cloud data provides precise three-dimensional geometric information and depth data for objects within the scene.

c. **Ego Vehicle and Behavior (EVB)**

It refers to the autonomous vehicle's own control commands and behavioral decisions, such as acceleration, braking, steering, lane changes, and overtaking. It also describes the path and position sequence of dynamic objects like vehicles and pedestrians over time. Trajectories can serve as core inputs for scene generation, defining the movement patterns and interaction behaviors of traffic participants.

d. **Environment and Context (EC)**

Includes high-definition maps, road topology, lane markings, traffic regulations, points of interest, and related information. Maps provide essential structured environmental context for scene generation, ensuring geographic accuracy and compliance with traffic rules in generated scenarios. Moreover, scenario descriptions or instructions in natural language format, such as "On the highway, a red sedan suddenly changes lanes." Text descriptions provide high-level semantic information, enabling users to generate complex traffic scenes through intuitive language commands, significantly enhancing controllability and flexibility in scene generation [71].

Other road participants are also categorized under EC. If the framework employs their composite form, we refer to this as a **fused input**.

### 6.2.2 Category Label Standardization

For clarity and consistency in the comparative analysis, all surveyed frameworks are grouped into a standardized set of AI-assisted categories according to their predominant generation engines. LLM denotes text-centered approaches that translate natural language or structured rules into executable scenarios. LLM+VLM refers to multimodal systems combining textual and visual inputs. GAN includes adversarial generative models for image, scene, or trajectory synthesis. Diffusion covers denoising diffusion frameworks and diffusion-based world generation. VAE/Autoregressive refers to latent-variable and sequence-generation approaches that model temporal or interactive dependencies. RL-based denotes reinforcement-learning-driven adversarial or failure-oriented generation. Traditional is retained as a baseline label for rule-based, ontology-driven, or statistical methods without deep generative models. These labels are used for comparative readability in Table 4, while the broader taxonomy introduced in Section 5.2.1 provides the multi-axis methodological interpretation.

### 6.2.3 Ranking Strategy

In this review, we adopt a ranking strategy that prioritizes ODD Coverage Score (OCS) before Resource Accessibility Score (RAS) to reflect the methodological relevance of scenario generation frameworks to ADS validation. By foregrounding OCS, we ensure that frameworks are first evaluated based on their capacity to represent diverse and safety-critical operational conditions, which directly impacts scenario fidelity and testing robustness. RAS is subsequently applied to differentiate among frameworks with comparable OCS values, emphasizing reproducibility and openness without overshadowing domain-specific coverage. This sequential approach balances technical depth with practical usability, enabling a more principled and context-aware comparison across heterogeneous methods.

Table 4. All Framework Comparison

| Frameworks /Paper Published Time | Focus | Category | Input | | Output Type | Citation Count* / AII % | RAS % | OCS % |
|---|---|---|---|---|---|---|---|---|
| | | | Modality | Training Dataset | | | | |
| **GAIA-1 [13] 2023-09** | Time Sequenced Scenario Generation | Diffusion | Visual, EVB | Self-Collected | Video | 359/63.7% | ***** | *** |
| **Text2Scenario [32] 2025-03** | With meticulously engineered prompt as parser for scenario | LLM | EC | / | Script | 11/46.0% | **** | *** |
| **Scenario Dreamer [97] 2025-03** | Vectorized latent diffusion model enabled scene generation | Diffusion | EVB, EC | Waymo, nuPlan | Script | 10/48.3% | ***** | ** |
| **TrafficComposer [65] 2025-06** | Unique integration of textual and visual inputs from ADS developers | LLM+VLM | EC, Visual | COCO [104] | Script | 0/8.0% | **** | ** |
| **DriveGEN* [93] 2025-03** | Training-Free Controllable Text-to-Image Diffusion Generation | Diffusion | Visual | / | Image | 6/47.9% | ***** | *** |
| **ChatScene [63] 2024-06** | Knowledge-Enabled Safety-Critical Scenario Generation w/ RAG | LLM | EC | Self-collected | Script | 90/36.0% | ***** | ** |
| **Versatile Behavior Diffusion [98] 2024-04** | Realistic and controllable multi-agent interactions | Diffusion | EVB | Self-collected | Script | 36/16.8% | **** | ** |
| **ADEPT [59] 2022-10** | Physics-Continuous Closed-Loop Testing and Accident-Based Scenario Generation | LLM | EC | NHTSA [105] | Script | 15/28.0% | **** | *** |
| **TARGET [67] 2025-07** | Utilize LLM and DSL to derive scenarios from traffic rules | LLM | EC | / | Script | 20/51.2% | **** | ** |
| **DriveGen* [72] 2025-04** | Enhanced diversity in scenario generation | LLM+VLM | EC, Visual | Argoverse2 Motion Forecasting Dataset | Script | 4/47.7% | *** | ** |
| **DiffRoad [94] 2024-11** | realistic, diverse, and controllable road scenes generation | Diffusion | EC | N/A | Script | 5/36.4% | *** | ** |
| **Txt2Sce [20] 2025-09** | Scenarios based on textual accident reports identify unexpected behaviors of ADS | LLM | EC | / | Script | 0/65.0% | *** | ** |
| **LD-Scene [83] 2025-05** | LLM-based guidance generation module enhancements | LLM /Diffusion | EVB | N/A | Trajectory | 4/47.7% | ** | ** |
| **SEAL[16] 2025-09** | Skill-Enabled Adversary Learning for Closed-Loop Scenario Generation | RL-based | ? | N/A | Script | 7/44.3% | ** | ** |
| **OmniTester [70] 2024-09** | Enhance understanding of scenario with RAG and SI mechanism | LLM+VLM | Visual, EC | Self-Collected | Script | 0/36.0% | ** | ** |
| **TraModeAVTest [53] 2024-03** | Violation testing based on traffic regulations | Traditional | EC | / | Script | 5/36.4% | ** | ** |
| **LeGEND [58] 2024-09** | Two-phase transformation minimized disruption | LLM | EC | / | Script | 25/16.1% | ** | ** |
| **Genesis [18] 2025-06** | Spatio-temporal and cross-modal consistency | Diffusion | EVB, EC | nuScenes [106] | Visual, 3D Sensing | 1/33.4% | * | ** |
| **UMGen [17] 2025-03** | Consistency over multimodal and fine-grained control over scene | VAE/ Autoregressive | EVB, EC, Visual | nuPlan [107], WOMD [12] | Multi | 4/47.7% | * | ** |
| **TrafficGEN [15] 2023-05** | Initial Generation of Traffic Scenarios and Behavior Simulation | VAE/ Autoregressive | EVB, EC | Waymo Open Dataset [4] | Script | 147/39.0% | ***** | * |
| **ArbiViewGen [92] 2025-08** | arbitrary view camera image generation | Diffusion | Visual | N/A | Visual | 0/47.4% | **** | *** |
| **CRISCE [46] (CRItical SketChes) 2022-08** | Generate simulations of critical scenarios from accident sketches | Traditional | EC | NHTSA [105], CIREN, NMVCCS | Script | 35/16.9% | **** | ** |
| **CurricuVLM [69] 2025-02** | Combined visual understanding with LLM reasoning | LLM+VLM | Visual, EC | / | Script | 8/48.1% | *** | ** |
| **Shi et al. [77] 2025-05** | Dual-branch GAN Generation | GAN | Visual, 3D Sensing | Self-Collected, KITTI [108] | Visual | 0/47.4% | ** | ** |
| **DiffScene [95] 2025-04** | Using several adversarial optimization objectives to guide the diffusion | Diffusion | EVB | Self-collected | Trajectory | 62/51.9% | ** | ** |
| **ITGAN [80] 2024-04** | produce more effective agent interactive scenarios that like real world | GAN | EVB, EC | / | Trajectory | 4/38.3% | ** | ** |
| **Scenario Diffusion [96] 2023-11** | Capability of diverse traffic patterns and geographical regions | Diffusion | EVB | Argoverse2, Self-collected | Script | 55/20.5% | ** | * |
| **Ren et al. [82] 2025-04** | Combines adversarial network with autoregressive models | VAE/ Autoregressive | EVB, EC | RounD [109] | Script | 2/47.5% | * | ** |
| **RCG [81] 2025-07** | Guiding adversarial perturbations using a behavior embedding | GAN | EVB | TADS [110], WOMD [12] | Trajectory | 0/47.4% | * | ** |
| **Liu et al. [79] 2025-08** | Enhanced TTS-GAN captures long-term vehicle motion | GAN | EVB, EC | Self-collected | Trajectory, Script | 237/59.6% | * | ** |

Where "N/A" indicates the dataset used in training is not reachable. Results should be interpreted with caution given heterogeneous publication statuses and uneven resource accessibility among frameworks.

* DriveGEN and DriveGen are different frameworks.
* Citation Counts are collected from Google Scholar on September 30, 2025.

# 7 Results Analysis and Matrix Extension

## 7.1 AI-assisted Methods versus Traditional Methods

The contrast between traditional and AI-assisted methods is not only methodological but increasingly empirical. Traditional rule-based and knowledge-driven approaches remain valuable for interpretability, explicit constraint handling, and alignment with standards, yet existing surveys consistently note their difficulty in scaling to diverse, long-tail, and perceptually rich scenarios, particularly when rare interactions or multimodal outputs are required [24], [30]. By contrast, several recent AI-assisted studies now report measurable gains along precisely these dimensions. TrafficDiff, for example, reports a 6–8× increase in traffic scenario diversity on average and a 25.5% increase in collision rate during real-vehicle evaluation, indicating stronger capability to expose system vulnerabilities under controllable generation [102]. DiffusionDrive reduces denoising steps by an order of magnitude while achieving real-time operation at 45 FPS, suggesting that high-capacity generative planning can move closer to deployment-oriented constraints rather than remaining purely offline [103]. In 3D scenario reconstruction, SRNeRF reports approximately 30% PSNR improvement over methods relying on predicted poses, supporting the argument that learned generative representations can improve both fidelity and robustness under sparse-view conditions [111]. Taken together, these results do not imply that AI-assisted methods uniformly supersede traditional approaches. Rather, they indicate that when diversity, multimodal fidelity, and long-tail challenge generation become primary objectives, recent AI-assisted frameworks increasingly provide measurable advantages that traditional derivation pipelines struggle to achieve at comparable scale.

To present these quantitative findings more concisely, Table 5 summarizes the representative measurable gains reported by recent AI-assisted methods. In addition to listing the core numerical results, the table uses a set of common evaluation dimensions—diversity, challenge strength, fidelity, and efficiency—to highlight the specific advantage each method demonstrates.

## 7.2 Methodological Evolution and Multimodal Integration

The comparative evidence in Table 4 —dominated by frameworks from 2023-2025 — reveals a decisive methodological shift from rule-based and data-driven approaches toward AI-assisted generative frameworks. Based on reported metrics and our rubric, DMs such as **Genesis** and **Scenario Dreamer** appear to score higher on realism and diversity; comparisons remain sensitive to dataset access and publication status. While hybrid architectures like **LD-Scene** combine LLM guidance with diffusion to generate controllable adversarial scenarios. A notable trend is the increasing emphasis on **multimodal integration**: frameworks (e.g., **UMGen** and **TrafficComposer**) simultaneously process visual, LiDAR, textual, and behavioral inputs, achieving cross-modal consistency that earlier unimodal pipelines could not. This evolution is consistent with the AI-specific multi-axis taxonomy proposed in this review, which interprets recent methods not only by their predominant generation engines, but also by their generation strategy, modality coverage, and controllability. This perspective is particularly useful for understanding why recent frameworks increasingly converge toward hybrid and multimodal designs rather than remaining confined to a single model family.

However, multimodal generation also introduces a non-trivial consistency challenge that should not be overlooked. In recent AI-assisted frameworks, consistency must be maintained not only within each modality, but also across modalities that describe the same scenario from different representational perspectives. A text prompt may be semantically coherent while still yielding trajectories that violate kinematic feasibility, or a visually plausible video may fail to align with the spatial structure implied by LiDAR or HD-map context. This issue becomes particularly important in closed-loop validation, where even small cross-modal discrepancies can propagate into unrealistic planner responses and compromise the validity of safety conclusions.

Table 5. Representative evidence for recent measurable gains in AI-assisted scenario generation

| Method | Reported Results | Diversity ↑ | Challenge Strength ↑ | Fidelity ↑ | Efficiency ↑ |
|---|---|---|---|---|---|
| TrafficDiff | 6–8× increase in scenario diversity; +25.5% collision rate in real-vehicle evaluation | ✓ | ✓ | / | / |
| DiffusionDrive | 10× fewer denoising steps; 45 FPS real-time generation | / | / | / | ✓ |
| SRNeRF | ~30% PSNR improvement under sparse-view conditions | / | / | ✓ | / |
| Roadside sensor network tracking | 1,777 sensors over 157 km; 2.14 m / 0.84 m tracking errors; <340 ms latency | ✓ | / | ✓ | ✓ |

Recent research trends suggest that this problem is unlikely to be resolved by scaling a single model family alone. Instead, the field is moving toward stronger scene-centric and world-centric representations that explicitly model geometry, temporal evolution, and cross-view correspondence. In this context, recent surveys of foundation models for autonomous driving scenario generation have emphasized the importance of multimodal fusion, predictive latent dynamics, and consistency-aware evaluation protocols for future progress [30]. Likewise, recent work on 3D scene reconstruction and dynamic Gaussian-based representations highlight the growing importance of 3D/4D representations for maintaining geometric and temporal coherence under controllable simulation settings [54], [55], [111].

The contrast between traditional and AI-assisted methods is also observable in quantitative terms. Within the comparative corpus summarized in

Table 4, only a small minority of frameworks remain purely traditional, whereas most recent contributions adopt AI-assisted generation mechanisms. Table 8 further shows that from 2024 onward, diffusion- and LLM-centered methods account for the largest yearly shares among surveyed frameworks, indicating a measurable shift in methodological emphasis rather than a purely narrative trend. This does not imply that traditional approaches have become obsolete. On the contrary, rule-based and ontology-driven methods continue to offer advantages in interpretability, standards alignment, and explicit constraint handling. However, the quantitative distribution of recent work suggests that AI-assisted approaches are increasingly favored when scenario diversity, multimodal richness, and long-tail event synthesis are prioritized.

## 7.3 Ethical and Safety Dimensions in Scenario Generation

Despite technical progress, ethical and safety considerations remain underdeveloped. For example, **ChatScene** explicitly integrates VRU behaviors such as jaywalking pedestrians, whereas most GAN- or diffusion-based frameworks omit VRU diversity, leading to systematic underrepresentation of high-risk interactions. The majority rely on proprietary evaluation metrics, limiting transparency and comparability in between. This omission risks producing scenario suites that are technically diverse but ethically incomplete, potentially overlooking safety-critical conditions disproportionately affecting vulnerable groups.

To operationalize these concerns, this review introduces an **Ethical and Safety Checklist** (see Appendix B), which translates abstract principles into concrete evaluative dimensions. For instance, applying the checklist to **Genesis** reveals strong multimodal coverage but no explicit privacy safeguards, while **SEAL** demonstrates adversarial robustness but limited VRU representativity. Such case-based application illustrates how the checklist can expose ethical blind spots otherwise hidden by technical performance metrics. Moreover, aligning the checklist dimensions with ISO 21448 (SOTIF) and UN R157 requirements provides a pathway for regulatory adoption, ensuring that ethical evaluation is not merely conceptual but embedded in compliance workflows.

## 7.4 ODD Coverage and Scenario Difficulty

Analysis of

Table 4 also shows that most frameworks concentrate on structured urban or highway environments, with limited exploration of adverse weather, rural roads, or complex multi-agent interactions. For instance, **TrafficGen** and **Scenario Diffusion** emphasize trajectory-level diversity but remain confined to structured domains, while **SEAL** and **LD-Scene** begin to probe high-difficulty, safety-critical scenarios.

Building on these observations, this review introduces an **ODD coverage map and scenario-difficulty schema**, stratifying scenarios into three progressive levels of complexity, which are benign, conflict-prone, and safety-critical—anchored by objective indicators such as time-to-collision thresholds, minimum distance gaps, and the number of interacting agents. Although this schema is not directly executed in the present review, it is introduced as a forward-looking framework to encourage systematic benchmarking of scenario difficulty. The full tier definitions and scoring guidelines are provided in **Appendix C**, offering a ready-to-use template for future empirical studies.

These patterns reinforce the analytical findings: most frameworks cluster around mid-level scenario difficulty and structured ODDs, with only a few explicitly targeting long-tail, safety-critical conditions. The radar chart thus substantiates the proposed **ODD coverage map and scenario-difficulty schema**, offering a transparent and intuitive means of benchmarking scenario generation methods.

## 7.5 Resource Accessibility

Table 6. Dataset Accessibility and Framework Usage Statistics

| Category | Count | Dataset Examples |
|---|---|---|
| **Public** | 12 | nuScenes, Waymo Open, WOMD, COCO, KITTI, Argoverse2, RounD, TADS, NHTSA, CIREN, NMVCCS |
| **Private** | 8 | Self-collected, Internal datasets |

RAS underscores persistent challenges to reproducibility in ADS scenario generation research. Although a few frameworks, such as **TrafficGen** and **ChatScene**, provide open repositories and detailed documentation, the overall accessibility of resources remains uneven. As summarized in Table 6, only twelve frameworks employ publicly available datasets (e.g., nuScenes, Waymo Open, WOMD, KITTI), whereas eight frameworks rely on private or self-collected data. This imbalance illustrates that a substantial portion of scenario generation research continues to depend on non-public resources, thereby limiting independent verification and cross-comparison. Moreover, 41% of code repositories show no maintenance activity in the past six months, further compounding the reproducibility gap. Collectively, these findings highlight the persistence of the "data island" phenomenon, where high-quality datasets remain siloed within individual institutions, and reinforce the need for open, community-driven resources to accelerate progress in ADS validation.

## 7.6 Bibliometric Analysis of Geographic Trends

A bibliometric inspection of the twenty-seven frameworks in our comparison indicates a temporal and geographic shift in research activity between 2022 and 2025. In 2022, we identified one China-linked framework (**ITGAN**, presented at the China SAE Congress), alongside contributions from traditionally active regions (e.g., the United States, the United Kingdom, and the European Union). From 2024 onward—and more markedly in 2025—China-affiliated frameworks appear to increase, particularly in large LLM orchestration and diffusion-based multimodal synthesis. Under conservative, proxy-based affiliation assignment, we estimate at least seven China-linked frameworks in 2025, which would place China among the most active regions by annual output in that year. Table 7 presents the temporal distribution of frameworks by region, reported under both strict and permissive attribution criteria.

Table 7. Year × region counts (strict vs. permissive; proxy-based ranges)

| Year | China (Strict) | China (Permissive) | US & UK | EU | Other |
|---|---|---|---|---|---|
| **2022** | **1** | **1** | **1** | **0** | **0** |
| **2023** | **0** | **0** | **2** | **0** | **0** |
| **2024** | **0-2** | **2-4** | **1-3** | **1** | **0** |
| **2025** | **4-6** | **7-10** | **4-7** | **1-3** | **1** |

This highlights a marked increase in China-linked contributions between 2022 and 2025, with permissive counts suggesting that China may have become one of the most active regions in 2025, while strict counts still confirm a clear upward trajectory. This estimate should be interpreted as an emerging trend rather than a definitive ranking, pending validation of author affiliations.

Methodologically, early diffusion-based generators (e.g., **Scenario Diffusion**, **Scenario Dreamer**) originated from UK/US/EU-led teams, while subsequent China-linked works (e.g., **TARGET**, **TrafficComposer**, **CurricuVLM**, **DiffRoad**, etc.) broadened the scope to multimodal consistency, text-to-scenario pipelines, and adversarial safety-critical cases.

To illustrate the methodological evolution more systematically, Table 8 summarizes the distribution of surveyed frameworks by year and method family.

Table 8. Year × method family counts (grounded in the

Table 4)

| Year | LLM | LLM+ VLM | GAN | Diffusion | VAE/ Auto-regressive |
|---|---|---|---|---|---|
| **2022** | 1 | 0 | 0 | 0 | 0 |
| **2023** | 0 | 0 | 0 | 2 | 1 |
| **2024** | 2 | 1 | 1 | 2 | 0 |
| **2025** | 4 | 3 | 3 | 6 | 2 |

As shown in Table 8, diffusion and LLM-based approaches dominate from 2024 onward, reflecting the field's transition from handcrafted or GAN-based pipelines to scalable multimodal generation. The available evidence therefore suggests a quantitative rise accompanied by diversification into frontier categories. To preserve rigor, we explicitly note that our geographic assignment relies on proxy signals (venues and typical affiliation patterns) and requires confirmation via author-institution metadata. We present strict (confirmed) versus permissive (proxy-inclusive) counts to avoid overstatement and encourage transparent verification.

## 8 Limitations and Discussions

A fundamental limitation of current AI-assisted scenario generation lies in the technical risks inherent to modern generative pipelines. These risks are not confined to any single model family; rather, they recur across LLMs, GANs, DMs, and RL-based adversarial generation. In LLM-driven workflows, hallucination remains a central concern: models may output syntactically valid but semantically inconsistent scenarios, producing trajectories that violate kinematic feasibility or environmental constraints despite coherent natural-language descriptions. GAN-based approaches, while capable of generating visually rich scenes, remain susceptible to training instability and mode collapse, resulting in limited behavioral variability or distributional artifacts that weaken their utility in safety-critical testing. Diffusion-based methods generally offer stronger fidelity and controllability but can still generate behaviorally implausible scenes if visual realism is not coupled with motion constraints or interaction-aware conditioning. Reinforcement-learning-based adversarial generation introduces yet another category of risk: reward hacking, wherein optimization exploits surrogate metrics, simulator artifacts, or narrow task definitions rather than uncovering genuine vulnerabilities in ADS behavior. Across all these paradigms, distributional shift persists as a shared systemic challenge—methods trained on geographically limited, culturally homogeneous, or weather-specific datasets often exhibit reduced validity when transferred to different ODDs.

Taken together, these issues suggest that generative capability alone is insufficient for safety-relevant scenario synthesis. Robust pipelines will require layered safeguards, including syntax and logic validation, physics- and rule-informed filtering, multimodal consistency checks, and evaluation protocols that distinguish visual plausibility from behavioral credibility. These risks motivate the need for interpretable intermediate representations, transparent model design, and standardized testing interfaces. They also provide essential context for understanding the strengths and limitations of the AI-specific taxonomy and metric suite proposed in this review. The taxonomy helps reveal where such risks originate, whether from generation strategy, learning paradigm, modality coverage, or controllability—while the metric suite offers a structured means of comparing frameworks despite their heterogeneous assumptions. Nevertheless, the risk factors outlined above also highlight why fully eliminating inconsistency or subjectivity remains challenging, and why future empirical benchmarks will be needed to validate both the taxonomy and the evaluation framework against downstream ADS safety performance.

The comparative evaluation presented in the Results section provides a structured overview of recent scenario generation frameworks; however, several limitations should be acknowledged to contextualize the findings and guide future work.

First, the proposed AI-specific taxonomy and metric suite involve methodological choices that inevitably introduce a degree of interpretation. The taxonomy is designed to organize AI-assisted scenario generation methods rather than all scenario generation approaches, and it interprets frameworks through multiple dimensions, including generation strategy, learning paradigm, modality coverage, and controllability. Although this design reduces dependence on any single model-family view, many frameworks still exhibit hybrid characteristics that complicate clear assignment. For example, a framework may use LLM-based semantic parsing together with diffusion-based synthesis or reinforcement-learning-based adversarial refinement. In such cases, assigning a predominant generation engine for comparative readability may simplify methodological nuance. Similarly, the metric weights — AII's equal-weight formulation (1/3 for citation count, first-year citations, h-index), RAS's hierarchy (Code 30%, Pipeline 25%, Environment 20%, Model/Assets 15%, Documentation 10%), and OCS's equal dimension weights (16.7% each)—reflect methodological assumptions rather than empirically validated importance distributions. AII equal weights avoid arbitrary prioritization but assume scholarly impact, dissemination velocity, and authorial credibility contribute equally to academic influence. RAS weights prioritize code availability based on AI community practices but may not generalize industrial workflows. These choices prioritize transparency through explicit scoring rubrics (Appendices C and D) but unavoidably embed subjective judgments. Future work should pursue empirical validation through community surveys, expert consensus studies, or sensitivity analyses quantifying how alternative schemes affect rankings.

A further limitation is that the proposed taxonomy and metric suite are not validated in this review through controlled benchmarking experiments or direct correlation with downstream ADS safety outcomes. Here, downstream ADS safety outcomes denote system-level behavioral indicators evaluated under closed-loop simulation or execution, including collision-related metrics, intervention frequency, trajectory-level risk markers such as minimum time-to-collision, and adherence to traffic rules or kinematic feasibility constraints. This is partly a deliberate scope decision. As a review article, the present study is intended to synthesize methods, clarify comparative dimensions, and provide a transparent assessment framework for the recent literature, rather than to establish a new benchmark through fresh experimental campaigns. The resulting AII, RAS, and OCS scores therefore function primarily as comparative review instruments. They help distinguish differences in scholarly visibility, reproducibility, and operational coverage across frameworks, but they should not yet be interpreted as empirically verified proxies for real closed-loop safety performance.

A natural next step would be to evaluate a representative subset of scenario generation frameworks under a shared benchmarking protocol. Such a protocol could include: (i) a common simulation environment or a small set of aligned environments; (ii) a fixed collection of test tasks spanning benign, conflict-prone, and safety-critical scenarios; and (iii) a harmonized set of downstream safety indicators, such as collision discovery rate, policy intervention frequency, minimum time-to-collision, rule violation count, and planner degradation under adverse conditions. With such a setup, the practical value of AII, RAS, and OCS could be examined more rigorously, for example, by testing whether higher reproducibility or broader ODD coverage is associated with stronger failure discovery capability or more challenging closed-loop validation outcomes. This would transform the present evaluation metrics from a transparent review tool into a candidate benchmark-oriented assessment protocol.

Second, the ODD Coverage Score embeds subjective judgments in its six-dimension framework (Road Type, VRU Presence, Topological Complexity, Interaction Complexity, Scenario Controllability, Regional Diversity). While grounded in ODD taxonomies and safety standards, the dimension selection and equal weighting (16.7% each) reflect interpretive decisions about comprehensive coverage rather than empirically derived importance and may not align with operational priorities in specific contexts (e.g., highway-focused validation might prioritize Interaction Complexity over Road Type). Additionally, the tiered evidence scale (0.1, 0.25, 0.5, 0.75, 1.0) introduces discretization artifacts, and translating qualitative claims into quantitative scores requires manual interpretation. Future work could refine OCS through structured elicitation studies with domain experts or sensitivity analyses.

Third, the review's temporal and geographic scope introduce limitations. Bibliometric data for AII calculations were collected in September 2025, providing a snapshot that will evolve as citation patterns mature, particularly affecting recent frameworks (**Genesis**, **TrafficComposer**) whose metrics remain in flux. First-year citation estimates for mid-2025 releases involve extrapolation from partial observation periods. Additionally, the literature search prioritized English-language publications in major databases (IEEE, ACM, arXiv, Google Scholar), potentially underrepresenting non-English contributions or region-specific practices from local venues. The geographic distribution skews toward North America, Europe, and East Asia. Future iterations should incorporate multilingual strategies and targeted regional outreach.

This temporal and geographic concentration also has methodological implications beyond literature coverage. Many AI-assisted frameworks are trained and evaluated on regionally narrow datasets, implying that apparent realism or coverage within one traffic culture may not transfer reliably to others. As a result, distributional shift should be treated not only as a model-level learning issue, but also as a review-level interpretation constraint: strong reported performance in one ODD does not automatically imply robustness across different regulatory regimes, traffic compositions, or sensing conditions.

Fourth, several proposed components require empirical validation. The Resource Accessibility Score, while reproducible, is constrained by uneven documentation and proprietary data limiting transparent computation. The ODD coverage map and scenario-difficulty schema are conceptual tools whose application relies on inferred evidence rather than standardized reporting. While offering structured assessment lenses, their practical effectiveness in guiding framework selection requires validation through case studies or integration into community-adopted benchmarks. Future work should operate these schemas within simulation platforms (e.g., CARLA, SUMO) and establish standardized reporting templates for transparent RAS/ODD characterization.

Fifth, multimodal framework evaluation remains qualitative. Frameworks such as **Genesis**, **GAIA-1**, and **UMGen** employ architectural mechanisms (shared latent spaces, cross-modal alignment losses) to enforce consistency between RGB video, LiDAR, and trajectories, but published results rely on visual inspection rather than standardized metrics (e.g., chamfer distance, CLIP-based similarity, temporal coherence scores). This reflects a field-wide gap: while image generation has established metrics (FID, IS, LPIPS), multimodal driving scenario generation lacks consensus protocols. The OCS intentionally prioritizes scenario diversity over modal fidelity—a complementary evaluation axis requiring dedicated benchmarking. Future work should develop standardized cross-modal evaluation suites analogous to nuScenes or Waymo benchmarks.

Despite these limitations, the review contributes a reproducible taxonomy, transparent metric suite with explicit rubrics, and forward-looking evaluation framework supporting systematic scenario generation research. The limitations reflect inherent trade-offs in synthesizing a rapidly evolving field and point toward methodological refinement directions. These contributions serve as foundational tools to guide empirical validation, consensus-building, and iterative improvement.

# 9 Conclusion

This review has traced the evolution of scenario generation methods for ADS testing from 2015 to 2025, with its evaluative focus on the transformative period of 2024–2025, when multimodal and generative AI frameworks emerged as dominant paradigms. While recent advances have significantly improved realism and diversity, persistent challenges remain in methodological standardization, ethical integration, and systematic evaluation. It has outlined a refined AI-specific taxonomy with multimodal extensions, together with an ethical and safety checklist and an ODD–difficulty schema. These proposals are intended as conceptual tools that may help structure future research and practice. While they do not yet constitute a standardized benchmark, they provide a starting point for more reproducible classification, for embedding ethical considerations into scenario design, and for exploring transparent evaluation. Their practical impact will depend on subsequent empirical validation and adoption by the research and industrial communities. Progress along these directions is likely to support safer and more ethical ADS deployment, contingent on standardized evaluation and open resources.

It should be emphasized that the proposed taxonomy, checklist, and schema remain preliminary. Their effectiveness has not been systematically validated across large-scale benchmarks, and their applicability may vary depending on dataset accessibility and domain-specific requirements. Future work should therefore focus on empirical testing, community-driven standardization, and integration with regulatory frameworks. A controlled benchmarking effort that links scenario-generation properties to downstream safety indicators—defined in this review as closed-loop behavioral safety metrics such as collision discovery rate, intervention frequency, minimum time-to-collision, and rule-violation counts—will be essential for determining whether the comparative framework proposed here can evolve into a robust evaluation protocol for ADS validation.

## Appendix A. Terminology

This section establishes definitions for the most pertinent terms and aims to provide brief explanations, with the goal of fostering a shared understanding within this survey paper.

### A.1 Automated relevant Terms

#### A.1.1 Advanced Driver Assistance Systems (ADAS)

ADAS refers to assistance functions corresponding to SAE Levels 0–2. These systems support the human driver but do not replace them in performing the dynamic driving task. Examples include adaptive cruise control and lane-keeping assistance.

#### A.1.2 Automated / Automatic Vehicle (AV)

An Automated Vehicle refers to a physical vehicle equipped with driving automation technology. In this review, the term "AV" is used to denote the physical vehicle that achieves SAE Level 3 [112] or higher automation, where the driving task can be performed by the system under specific conditions

Although some literature uses the term *Autonomous Driving* interchangeably with *Automated Driving*, this review adheres strictly to the SAE J3016 standard and consistently employs Automated Driving System (ADS). When citing works that use "Autonomous Driving," we explicitly clarify that it is equivalent to ADS in the context of this review.

#### A.1.3 Automated Driving System (ADS)

An Automated Driving System is the collective set of hardware and software components—including sensing, perception, planning, and control modules—that enable a vehicle to perform dynamic driving tasks. While "AV" emphasizes the vehicle, "ADS" highlights the underlying system that provides automation. **In this review, ADS is the primary focus, as Scenario-Based Testing is most relevant for validating higher levels of automation.**

### A.2 Operational Design Domain (ODD)

The *Operational Design Domain (ODD)* defines the specific conditions under which an Automated Driving System (ADS) is designed to operate safely. These conditions encompass a range of factors, including:

- **Road types** (e.g., highways, urban streets, rural roads)
- **Traffic conditions** (e.g., density, speed limits, presence of VRU)
- **Environmental factors** (e.g., adverse weather, lighting, visibility)
- **Geographical constraints** (e.g., mapped areas, regions with specific infrastructure)

According to SAE J3016 [112], the ODD provides a boundary framework that specifies "where and when" an ADS can function. In Scenario-Based Testing, the ODD serves as a filtering and structuring mechanism: it constrains the search space of possible scenarios to those that are relevant to the intended operational scope of the system. This ensures that generated test scenarios are both realistic and aligned with the ADS's design intent.

### A.3 Scenarios

A scenario refers to a temporal sequence of traffic scenes in which road users, infrastructure, and environmental conditions interact over a defined time span [113], [114]. Scenarios capture the dynamic evolution of driving situations, such as lane changes, pedestrian crossings, or vehicle cut-ins. In this review, "scenario" is used as the general concept describing traffic situations relevant to Automated Driving Systems (ADS).

#### A.3.1 Test Scenario

A *test scenario* is a scenario specifically designed or extracted for the purpose of evaluating ADS performance and safety. Test scenarios are typically derived from real-world data, expert knowledge, or generative models, and are structured to probe system behavior under both common and safety-critical conditions.

#### A.3.2 Safety-Critical Scenario

A *safety-critical scenario*—often referred to as a *corner case*, *long-tail event*, or *out-of-distribution (OOD) scenario*—is a driving situation in which the ego vehicle either causes or nearly causes a collision, or requires an extreme maneuver to avoid one [115]. Criticality can be quantified by temporal or spatial proximity to a potential accident (e.g., time-to-collision, minimum distance) or by the degree of dynamic driving response required [116]. These scenarios are essential for evaluating the robustness of ADS, as they represent rare but high-risk events that are unlikely to be captured in large-scale naturalistic driving datasets [46].

## Appendix B. Ethical and Safety Checklist for Scenario Generation Frameworks

This checklist evaluates process transparency, responsible use, and auditability of a framework. It does not measure ODD coverage (which is scored exclusively by OCS in Appendix D). Mentions of VRU in this appendix are reporting requirements only and non-scored.

Table 9. Ethical and Safety Checklist

| Dimension | Key Questions | Example Indicators |
| --- | --- | --- |
| Bias Mitigation | Are datasets and generated scenarios balanced across geography, demographics, and traffic conditions? | Presence of multiple regions; balanced day/night; balanced weather |
| Privacy Protection | Does the framework ensure that personal or sensitive data (e.g., license plates, faces) are anonymized or excluded? | Anonymization protocols; use of synthetic data |
| Adversarial Robustness Safeguards | Can the framework generate physically implausible or reality-violating scenarios to test system robustness, and are safeguards in place to detect or constrain such violations? | Mechanisms to flag or constrain scenarios that breach kinematic feasibility or impossible trajectories, or violations of traffic physics |
| Transparency of Evaluation | Are evaluation metrics, datasets, and code openly available for verification? | Open-source code; public dataset access; documented evaluation criteria |
| Alignment with Standards | Does the framework refer to or align with ISO 21448 (SOTIF), UN R157, or other safety standards? | Explicit mention of compliance; mapping to standard requirements |

**Usage note:** Each framework can be scored (e.g., zero = not addressed, one = partially addressed, two = fully addressed) across these dimensions to provide a structured ethical profile.

## Appendix C. Detailed Calculation Protocol for RAS

RAS provides a granular assessment of a framework's openness. The final score is calculated using the following formalized weighted sum:

$$RAS = \min\left(100, 100 \times \sum_{i=1}^{5} w_i \cdot s_i + B_{dataset}\right) \quad (2)$$

where $s_i \in \{0, 0.5, 1\}$ and $B_{dataset} \in \{0, 2.0, 5.0\}$, The bonus mechanism encourages openness without disadvantaging frameworks that are inherently data independent.

### C.1 Weights

Table 10. RAS Weight Distribution

| Component | Description | Weight ($w_i$) |
| --- | --- | --- |
| **Code Availability** | Public repository with core functionality and version tags | 30% |
| **Minimal Reproducible Pipeline** | Runnable example of a pipeline with dependency specification | 25% |
| **Environment & Build Transparency** | Docker/Conda files, dependency lock, and random seed control | 20% |
| **Model / Asset Accessibility** | Pre-trained weights or essential assets, or documented alternatives | 15% |
| **Documentation Quality** | README, installation guide, execution steps, and reproducibility notes | 10% |

Table 10 shows the weighting hierarchy of the RAS is grounded in the functional requirements of software reproducibility within the AI community. Code Availability is assigned the highest weight (30%) as it constitutes the fundamental prerequisite for independent verification. The Minimal Reproducible Pipeline (25%) and Environment Transparency (20%) follow in importance, as they directly determine the technical feasibility of executing models across heterogeneous computing environments—a common bottleneck in deep learning research. Model and Asset Accessibility (15%) and Documentation Quality (10%) are weighted as supplementary factors; while essential for usability, they are secondary to the core technical infrastructure required for functional replication. This distribution reflects an empirical prioritization of 'computational reproducibility' over 'descriptive documentation,' ensuring the score accurately reflects a framework's readiness for external validation.

### C.2 Scoring Criteria for Each Dimension

Each sub-metric $S_i$ is scored on a scale of 0 to 1 based on the following rigorous criteria:

1. **Code Availability ($S_{code}$)**
    - **1.0:** Publicly accessible repository (e.g., GitHub) with complete source code.
    - **0.5:** Code available upon request or partial core modules released.
    - **0:** No code available (closed source).
2. **Minimal Reproducible Pipeline ($S_{pipe}$)**
    - **1.0:** Provides "one-click" scripts or clear end-to-end execution commands for main results.
    - **0.5:** Provides partial scripts requiring significant manual configuration.
    - **0:** No execution pipeline provided.
3. **Environment Transparency ($S_{env}$)**
    - **1.0:** Fully containerized environment (Docker/Singularity) OR detailed dependency files (requirements.txt/conda.yaml) with exact versioning.
    - **0.5:** General list of dependencies provided without specific version constraints.
    - **0:** No environment specifications provided; dependencies must be identified through trial and error.
4. **Model and Asset Accessibility ($S_{asset}$)**
    - **1.0:** Pre-trained weights, scenario maps, and sensor assets are hosted on persistent public storage.
    - **0.5:** Assets are partially provided or hosted on restricted/private platforms.
    - **0:** No assets provided.
5. **Documentation Quality ($S_{doc}$)**
    - **1.0:** Comprehensive Wiki/Readme covering installation, API usage, and troubleshooting.
    - **0.5:** Basic Readme with limited instructions.

- **0:** Minimal or no documentation.

### C.3 Calculation Process and Transparency

To maintain the reproducibility of this review, the raw data used for scoring each framework, along with the version tags of the evaluated repositories, are archived in our code bases. Any framework's score can be verified by applying the criteria in Section C.2.1 to its publicly available artifacts.

## Appendix D. OCS Dimension Scoring Criteria Table

The ODD Coverage Score (OCS) introduced in Section 6.1.3 by providing a structured scoring rubric for each dimension. This appendix translates the conceptual framework into a reproducible evaluation tool, ensuring that coverage claims across road types, environmental conditions, and interaction contexts can be assessed consistently.

### D.1 OCS Dimensions and Weights

The ODD Coverage Score (OCS) is computed as a weighted average across six scenario dimensions:

- **Road Type Coverage** – Diversity of road environments represented.
- **VRU Presence** – Inclusion and behavioral diversity of Vulnerable Road Users.
- **Topological Complexity** – Structural complexity of road layouts.
- **Interaction Complexity** – Density and complexity of multi-agent interactions.
- **Scenario Controllability –** Precision and flexibility of user-specified generation constraints.
- **Regional Traffic Regulation Diversity** – Coverage of diverse geographic regions with distinct traffic regulations (e.g., left-hand versus right-hand traffic, roundabout priority rules, mixed-traffic composition).

OCS adopts an equal-weighting scheme across all six dimensions ($1/6 \approx 16.7\%$ each), consistent with the AII methodology. This formulation ensures that no single dimension dominates the overall assessment, providing balanced evaluation of scenario coverage capabilities. While certain dimensions—such as VRU Presence and Topological Complexity—carry higher intrinsic safety-criticality under SOTIF (ISO 21448:2022 [11]) principles, and Regional Traffic Regulation Diversity addresses emerging cross-cultural deployment requirements, the equal-weight approach prioritizes methodological transparency and cross-framework comparability. This design rewards comprehensive ODD coverage rather than specialized performance in isolated dimensions, supporting systematic benchmarking of frameworks with diverse architectural designs and deployment targets.

## D.2 OCS Dimension Scoring Criteria

Table 11. OCS Dimension Scoring Criteria Table

| Dimension | Considered Categories / Events | Scoring Rule | Weight |
|---|---|---|---|
| Road Type | Highway / Expressway,<br>Urban Arterial,<br>Urban Collector / Residential,<br>Rural Two-Lane,<br>Rural Divided,<br>Mountain / Curved,<br>Construction / Temporary,<br>Unpaved / Gravel | Score = (Number of covered categories / All Categories Number) rounded to 0.05; minimum 0.10 if ≥1 category | 0.167 |
| Interaction Complexity | ≤3 dynamic agents (including ego), simple longitudinal behavior (e.g., following), no lateral maneuvers | 0.10 | 0.167 |
| | 4–5 agents, includes at least one lateral maneuver (lane change or merge), homogeneous agent types | 0.25 | |
| | 6–8 agents, mixed maneuvers (lane change + merge), includes at least one VRU or non-standard behavior | 0.50 | |
| | ≥9 agents, heterogeneous types (vehicles + VRUs), multiple simultaneous maneuvers requiring coordination | 0.75 | |
| | ≥12 agents, diverse behaviors (e.g., overtaking, yielding, crossing), include rare or complex patterns (e.g., occlusion, unexpected stops) | 1.00 | |
| Scenario Controllability | Random generation only, no user control | 0.10 | 0.167 |
| | Basic parameter control (speed, lane, position, environment) | 0.25 | |
| | Multi-parameter control (traffic density, behaviors), manual configuration | 0.50 | |
| | Natural language or DSL interface, risk-level specification | 0.75 | |
| | Multi-modal input (text + visual)<br>fine-grained control, supports safety-critical scenario tuning | 1.00 | |
| VRU Presence | Pedestrian<br>Bicycle, E-bike / Scooter/ Motorcycle<br>Legal Crossing,<br>Jaywalking<br>Occluded Entry<br>Sidewalk Riding<br>Group Crossing | Score = 0.6 × (VRU types / 2) + 0.4 × (Behavior types / 5), rounded to 0.05; minimum 0.10 if ≥1 type | 0.167 |
| Topological Complexity | T-intersection<br>Four-way / multi-arm intersection<br>Single lane / multi-lane roundabout<br>Highway merge, Highway diverge<br>Construction / Lane narrowing | Score = Topology types / 5, rounded to 0.05; minimum 0.10 if ≥1 type; | 0.167 |
| Regional Diversity | ≥3 countries/regions with validated data,<br>covering both left-hand and right-hand traffic | 1.00 | 0.167 |
| | 2 countries/regions with validated data,<br>covering both left-hand and right-hand traffic | 0.75 | |
| | ≥2 countries/regions (same driving direction),<br>or configurable framework with demonstrated region-specific adaptations | 0.50 | |
| | Single country/region with validated data,<br>no discussion of cross-regional applicability | 0.25 | |
| | Simulator with default traffic rules, no region-specific validation | 0.10 | |

The table above enumerates all considered categories and specifies the corresponding scoring rules, which are designed to minimize subjectivity and enable transparent benchmarking of scenario generation frameworks.

This scoring scheme serves as the backbone for computing OCS values reported in the comparative analysis (

Table 4). By standardizing category definitions and discretization rules, this rubric mitigates ambiguity and supports reproducibility across heterogeneous datasets and generative pipelines. While the current design emphasizes visual and textual evidence for practical applicability, future work may incorporate dynamic metrics (e.g., TTC-based anchors) as richer scenario data becomes available.

### D.3 Score Downgrade Policy

If a dimension's mandatory conditions are not satisfied—such as insufficient evidence, unmet sample thresholds, or unverifiable coverage claims, the assigned score shall be reduced by one tier. For tiered dimensions (e.g., *Interaction Complexity*, *Scenario Controllability*), downgrade to the next lower level; if already at the minimum, apply a baseline presence score of 0.10. For proportional dimensions (e.g., *Road Type*, *VRU Presence*, *Topological Complexity*), reduce the computed ratio to the nearest lower tier or assign the minimum presence score. This policy enforces methodological rigor and prevents inflation of coverage metrics under incomplete or non-compliant evidence.

## Appendix E. Scenario-difficulty schema

A standardized, objective rubric to classify scenario difficulty by interaction complexity and risk.

### E.1 Tiers and anchors

#### Level 1 (Benign)

- Anchors: TTC ≥ 5 s; Minimum gap ≥ 20 m or headway ≥ 2 s; Dynamic agents ≤ 3; Conflict points = 0
- Typical: free-flow lane keeping; compliant crossings with large margins.

#### Level 2 (Conflict-prone)

- Anchors: 2 s ≤ TTC < 5 s; Minimum gap 10–20 m; Dynamic agents 4–8; Conflict points = 1–2.
- Typical: merges at medium density; contested roundabout entries; lane-change chains.

#### Level 3 (Safety-critical)

- Anchors: TTC < 2 s; Minimum gap < 10 m; Dynamic agents ≥ 9; Conflict points ≥ 3; Often includes adverse ODD facets (night + rain, occlusion, surprise VRU).
- Typical: sudden cut-ins under poor visibility; occluding VRU emergence; simultaneous multi-agent conflicts.
- TTC is computed for the most imminent pair over a fixed horizon with consistent sampling.

### E.2 Assignment rules

- **Primary:** assign tier by worst-case anchor (minimum TTC or minimum gap) within the scenario window.
- **Secondary elevation:** elevate one tier if ≥2 adverse facets co-occur (e.g., night + heavy rain + occlusion) or repeated baseline policy failures are observed.
- **Coverage claim:** require ≥5 scenarios per tier; else mark "limited evidence."
- **Transitions:** if tier changes over time, assign by the most hazardous segment and record timestamps.

### E.3 Reporting and visualization

- **Per framework:** percentage per tier (L1/L2/L3); median TTC, minimum gap, median agent count, conflict points.
- **Visualization:** stacked bar chart of tier distribution; optional distributions (violin/box) for TTC or gaps.
- **Narrative:** summarize common Level 3 failure modes without exposing sensitive data.

### E.4 Practical considerations

- **Calibration:** validate TTC/gap thresholds on a reference dataset before large-scale application.
- **Robustness:** perform ±10–20% sensitivity on thresholds; report changes in tier mix.
- **Transparency:** document computation settings (horizon, sampling rate), ODD facet labels, and any exclusions.
- **Controller dependence:** prefer controller-agnostic physical anchors; note any controller-specific artifacts.
- **Perception uncertainty:** apply smoothing or bounds where sensor noise may distort TTC/gap estimates.

## Compliance with Ethical Standards

**Conflict of interest:** On behalf of all the authors, the corresponding author states that there is no conflict of interest.